%% file: steady.tex
\input type
\ifdraft
    \documentclass[dvips]{jfm}      % save paper
\else
    \documentclass[dvips]{jfm}      % as it appears in journal
\fi
\input setupSteady
\input defSteady

\title[Geometry of plane Couette flow]
{Visualizing the geometry of state space in plane Couette flow}

\author[
J. F. Gibson,
J. Halcrow,
    and
P. Cvitanovi\'c
]{
J.\ns F.\ns G\ls I\ls B\ls S\ls O\ls N
,\ns
J.\ns  H\ls A\ls L\ls C\ls R\ls O\ls W
\ns
\and
P.\ns C\ls V\ls I\ls T\ls A\ls N\ls O\ls V\ls I\ls \'C
}

\affiliation{
School of Physics,
Georgia Institute of Technology,
Atlanta, GA  30332, USA 
}

\pubyear{2007}
\volume{638}
\pagerange{119--126}
\date{9 August 2007 and in revised form 11 February 2008 }

\begin{document}

\maketitle

\renewcommand{\eqb}{equilibrium}
\renewcommand{\Eqb}{equilibrium}
\renewcommand{\eqba}{equilibria}
\renewcommand{\Eqba}{Equilibria}
\renewcommand{\reqb}{relative equilibrium}
\renewcommand{\Reqb}{Relative equilibrium}
\renewcommand{\reqba}{relative equilibria}
\renewcommand{\Reqba}{Relative equilibria}

\input abstract

\section{Introduction}
\label{s:intro}
\input intro

\section{\PCf}
\label{s:review}
\input PCFreview

\subsection{Symmetries}
\label{s:PCFsymm}
\input PCFsymm

\section{Invariant solutions of plane Couette}
\label{s:equilibria}
\input equilibria

\section{The geometry of plane Couette \statesp}
\label{s:geometry}
\input geometry

\section{Conclusion and perspectives}
\label{s:conclusions}
\input conclusions

\begin{acknowledgments}
\input ackn
\end{acknowledgments}

\appendix

\section{Tabulation of numerical results}
\label{s:appeTables}
\input appeTables

\bibliographystyle{jfm}

\bibliography{steady}

\end{document}

%% file: type.tex
\newif\ifdraft
\draftfalse     % For final version, no comments

\newif\ifcolorfigs
\colorfigstrue     % For color copies

%% file: setupSteady.tex
\usepackage{ifpdf}
\ifpdf
  \usepackage[pdftex]{color,graphicx}
\else
  \usepackage[dvips]{graphicx}
\fi
\usepackage{natbib}
\usepackage{upmath}
\usepackage{amsmath,amsfonts,amssymb,amsbsy,amscd}
\usepackage{color}
\usepackage{url}
\usepackage{ifthen}

\usepackage[active]{srcltx}

\DeclareGraphicsExtensions{.pdf,.png,.gif,.jpg}
  \newcommand{\colorcomm}[2]{#1}
\providecommand\bnabla{\boldsymbol{\nabla}}

\newsavebox{\astrutbox}
\sbox{\astrutbox}{\rule[-5pt]{0pt}{20pt}}

%% file: defSteady.tex
    \ifdraft    % display comments in text
\newcommand{\Preliminary}[1]
             {\color{magenta}
              \marginpar{$\Downarrow${\color{magenta}\footnotesize PRELIMINARY}}
               #1
              \marginpar{$\Uparrow${\color{magenta}\footnotesize PRELIMINARY}}
              \color{black}
             }

\newcommand{\PC}[1]{$\footnotemark\footnotetext{PC: #1}$}

\newcommand{\FW}[1]{$\footnotemark\footnotetext{FW: #1}$}

\newcommand{\JFG}[1]{$\footnotemark\footnotetext{JG: #1}$}

\newcommand{\JH}[1]{$\footnotemark\footnotetext{JH: #1}$}

\newcommand{\file}[1]{$\footnotemark\footnotetext{{\bf file} #1}$}
\def\mycomment #1#2 {\noindent \textbf{\underline{#1}}: \emph{#2}}
    \else   % drop comments
\newcommand{\Preliminary}[1]{}

\newcommand{\PC}[1]{}
\newcommand{\JFG}[1]{}
\newcommand{\FW}[1]{}
\newcommand{\JH}[1]{}

\newcommand{\file}[1]{}
\def\mycomment #1#2 {}
    \fi
\newcommand{\refeq}  [1] {(\ref{#1})}

\newcommand{\reffig} [1] {figure~\ref{#1}}

\newcommand{\refFig} [1] {Figure~\ref{#1}}

\newcommand{\reftab} [1] {table~\ref{#1}}

\newcommand{\refsect}[1] {\S\,\ref{#1}}

\newcommand{\refSect}[1] {\S\,\ref{#1}}

\newcommand{\beq}{\begin{equation}}

\newcommand{\nnu}{\nonumber}
\newcommand{\eeq}{\end{equation}}
\newcommand{\ee}[1] {\label{#1} \end{equation}}

\newcommand{\bea}{\begin{eqnarray}}
\newcommand{\eea}{\end{eqnarray}}
\newcommand{\barr}{\begin{array}}
\newcommand{\earr}{\end{array}}
\def\ie{{\em i.e.\ }}
\newcommand{\NS}{Navier-Stokes}
\newcommand{\NSe}{Navier-Stokes equation}

\newcommand{\KSe}{Kuramoto-Sivashinsky equation}
\newcommand{\Reynolds}{\textit{Re}}  % Reynolds number
\newcommand{\pCf}{plane Couette flow}
\newcommand{\PCf}{Plane Couette flow}
\newcommand{\ubranch}{upper-branch}

\newcommand{\lbranch}{lower-branch}

\newcommand{\eqb}{equilibrium}
\newcommand{\Eqb}{equilibrium}
\newcommand{\eqba}{equilibria}
\newcommand{\Eqba}{Equilibria}

\newcommand{\reqb}{relative equilibrium}
\newcommand{\Reqb}{Relative equilibrium}
\newcommand{\reqba}{relative equilibria}
\newcommand{\Reqba}{Relative equilibria}
\newcommand{\po}{periodic orbit}

\newcommand{\rpo}{relative periodic orbit}

\newcommand{\cohStr}{coherent state}

\newcommand{\stateDsp}{state-space}

\newcommand{\statesp}{state space}
\newcommand{\ew}{eigen\-value}
\newcommand{\ev}{eigen\-vector}

\newcommand{\bu}{\ensuremath{{\bf u}}}
\newcommand{\bv}{\ensuremath{{\bf v}}}
\newcommand{\bff}{\ensuremath{{\bf f}}}

\newcommand{\be}{{\bf e}}
\newcommand{\bx}{{\bf x}}
\newcommand{\ex}{{\hat{\bf x}}} % unit vectors
\newcommand{\ey}{{\hat{\bf y}}}
\newcommand{\ez}{{\hat{\bf z}}}
\newcommand{\bPhi}{{\bf \Phi}}

\newcommand{\stagn}{\ensuremath{\text{\tiny EQ}}}% JFG equilib/stagnation point
\newcommand{\uTW}{\ensuremath{\bu_{\text{\tiny TW}}}}
\newcommand{\uEQ}{\ensuremath{\bu_{\text{\tiny EQ}}}}
\newcommand{\vEQ}{\ensuremath{\bv_{\text{\tiny EQ}}}}
\newcommand{\uLM}{\ensuremath{\bu_{\text{\tiny LM}}}}
\newcommand{\uLB}{\ensuremath{\bu_{\text{\tiny LB}}}}
\newcommand{\uNB}{\ensuremath{\bu_{\text{\tiny NB}}}}
\newcommand{\uUB}{\ensuremath{\bu_{\text{\tiny UB}}}}

\newcommand{\ssp}{a}                          % state space point

\newcommand{\ben}[1]{{\be}_{#1}}
\newcommand{\beUBg}[1]{\ensuremath{\be_{#1}}}
\newcommand{\beUBl}[1]{\ensuremath{\be_{#1}^{\lambda}}}
\newcommand{\sspn}[1]{a_{#1}}               % state space point
\newcommand{\sspg}[1]{\ssp_{#1}}        % state space point global
\newcommand{\sspl}[1]{\ssp_{#1}^{\lambda}}  % state space point local

\newcommand{\bbU}{\mathbb{U}}
\newcommand{\bbUsymm}{\bbU_{S}}

\newcommand{\pd}[2]{\frac{\partial #1}{\partial #2}}
\newcommand{\Norm}[1]{\|{#1}\|}
\newcommand{\grad}{\boldsymbol{\nabla}}

\newcommand{\timeAver} [1]{\overline{#1}}
\newcommand{\shift}{\ensuremath{\ell}}
\newcommand\period[1]{{T_{#1}}}     %continuous cycle period
\newcommand{\ucoeff}{u}             % spectral coefficient for high-d u expansion
\newcommand{\vField}{\ensuremath{{\bf F}}} % symbolic Navier-Stokes dynamics
\newcommand\Lyap{\lambda}                       %Lyapunov exponent

\newcommand{\eigExp}[1][]{
\ifthenelse{\equal{#1}{}}{\ensuremath{\lambda}}{\ensuremath{\lambda^{(#1)}}}
                        }
\newcommand{\eigRe}[1][]{
\ifthenelse{\equal{#1}{}}{\ensuremath{\mu}}{\ensuremath{\mu^{(#1)}}}
                        }
\newcommand{\eigIm}[1][]{
  \ifthenelse{\equal{#1}{}}{\ensuremath{\omega}}{\ensuremath{\omega^{(#1)}}}
            }

\newcommand{\tny}[1]{{\text{\tiny {#1}}}}

\newcommand{\Wmnfld}[2]{%
\ifthenelse{\equal{#2}{}}{\ensuremath{W_{#1}}\!}
                         {\ensuremath{W^{#1}_{\text{\tiny #2}}}\!} %Negative space is screwing up spacing in text
                        }

%% file: abstract.tex
% abstract.tex
% $Author: predrag $ $Date: 2008-02-09 23:50:45 +0530 (Sat, 09 Feb 2008) $
% based of J Fluid Mechanics example JFM2esam.tex,v2.0, 27th July 2004

Motivated by recent experimental and numerical studies of coherent
structures in wall-bounded shear flows, we initiate a systematic
exploration of the hierarchy of unstable invariant solutions of the
Navier-Stokes equations. We construct a dynamical, $10^5$-dimensional
state-space representation of plane Couette flow at $Re = 400$ in a
small, periodic cell and offer a new method of visualizing invariant
manifolds embedded in such high dimensions. We compute a new
equilibrium solution of plane Couette flow and the leading eigenvalues
and eigenfunctions of known equilibria at this $Re$ and cell size.
What emerges from global continuations of their unstable manifolds is
a surprisingly elegant dynamical-systems visualization
of moderate-$Re$ turbulence. The invariant manifolds tessellate the
region of state space explored by transiently turbulent dynamics with
a rigid web of continuous and discrete symmetry-induced heteroclinic
connections.

%% file: intro.tex
% intro.tex
% $Author: gibson $ $Date: 2008-02-11 00:38:43 +0530 (Mon, 11 Feb 2008) $

% \section{Introduction}

% Hopf
%

In a seminal paper, \cite{hopf48} envisioned the function space of
{\NS} velocity fields as an infinite-dimensional
\statesp, parameterized by viscosity, boundary conditions and
external forces, in which each $3D$ fluid velocity field is
represented as a single point. Laminar states correspond to {\eqba}
that are globally stable for sufficiently large viscosity. As the
viscosity decreases (Reynolds number increases), turbulence
sets in, represented by chaotic \stateDsp\ trajectories. Hopf's
observation that viscosity causes {\stateDsp} volumes to contract
under the action of dynamics led to his key conjecture: that
long-term, typically observed solutions of the \NSe s lie on
finite-dimensional manifolds embedded in the infinite-dimensional
{\statesp} of allowed velocity fields. These manifolds, known today as
`inertial manifolds,' are well-studied in the mathematics of
spatio-temporal PDEs. Their finite dimensionality for non-vanishing
viscosity parameters has been rigorously established in certain
settings by \cite{FNSTks85} and collaborators.

Since Hopf's time, engineers and applied mathematicians have assembled a
body of empirical evidence that moderately turbulent flows exhibit organized,
intrinsically low-dimensional behavior for a variety of conditions
(see \cite{Holmes96}, \cite{Pan97}, and \cite{Robinson91} for good overviews
of this large body of work). The experiments of \cite{KRSR67}, for example, revealed spatially
organized streaks in the turbulent boundary layer. The numerical simulations of
\cite{KMM87} opened access to the full $3D$ velocity field of channel flows
and paved the way for more detailed studies of organization in wall-bounded
flows. The work of \cite{HaKiWa95} began a very fruitful line of research;
it identified from numerical simulations a remarkably well-defined, quasi-cyclic
process among streamwise streaks and vortices (or `rolls') in low-Reynolds number {\pCf}.
\cite{W95a,W97} further developed these ideas into a
`self-sustaining process theory' that explains the quasi-cyclic roll-streak
behavior in terms of the forced response of streaks to rolls, growth of streak
instabilities, and nonlinear feedback from streak instabilities to rolls.

The preponderance of recurrent, coherent states in wall-bounded shear
flows suggests that their long-time dynamics lie on low-dimensional
\stateDsp\ attractors. This has motivated a number of researchers to model
such flows with low-dimensional dynamical systems.
\cite{Aubry88,Holmes96} used `Proper Orthogonal
Decomposition' [POD] of experimental data and Gal\"erkin projection of
the Navier-Stokes equations to produce low-order models of
coherent structures in boundary-layer turbulence. These models reproduce
some qualitative features of the boundary layer, but the quantitative
accuracy and the validity of simplifying assumptions in their
derivation are uncertain (\cite{Zhou92,Sirovich94,GibsonPhD}).
POD models for plane Couette were developed by \cite{SMH05}

Another class of low-order models of \pCf\ derives from the
`self-sustaining process theory' discussed above
(\cite{DV2000,MFE04,MFE04b,Mann04,Skuf05}). These models
use analytic basis functions explicitly designed to represent
the streaks, vortices, and instabilities of the self-sustaining
process, compared to numerical basis functions of the POD, which
represent statistical features of the flow. They improve on
the POD models by capturing the linear stability of the laminar
flow and saddle-node bifurcations of non-trivial 3D {\eqba}
consisting of rolls, streaks, and streak undulations. The work of
\cite{SYE05},
based on a \cite{Schmi99} 9-variable model, offers an elegant
dynamical systems picture, with the stable manifold of a periodic
orbit defining the basin boundary that separates the turbulent and
laminar attractors at $\Reynolds < 402$ and the stable set of a
higher-dimensional chaotic object defining the boundary at higher
$\Reynolds$. However, these models share with POD models a sensitive
dependence on modeling assumptions and uncertain quantitative relations
to true Navier-Stokes flows. A systematic study of the convergence of
POD/Gal\"erkin models of \pCf\ to fully-resolved simulations
indicates that dimensions typical in the literature ($10$-$10^2$)
are orders of magnitude too low for either short-term quantitative
prediction or reproduction of long-term statistics (\cite{GibsonPhD}).

The lack of quantitative success in low-dimensional modeling motivates yet
another approach: the calculation of {\em exact invariant solutions}
of the fully-resolved Navier-Stokes equations.
The idea here is to bypass low-dimensional modeling and to treat
fully-resolved CFD algorithms directly as very high-dimensional dynamical
systems. \cite{N90} computed a `lower-branch' and `upper-branch' pair of
nontrivial {\eqba} solutions to plane Couette flow by continuation and
bifurcation from a wavy vortex solution of Taylor-Couette flow. 
Starting with physical insights from the self-sustaining process theory,
\cite{W98,W01,W03} generated, {\it ab initio}, families of exact $3D$
{\eqba} and traveling waves of Navier-Stokes in plane Couette and Poiseuille
flows for a variety of boundary conditions and $\Reynolds$ numbers, using a
$10^4$-dimensional Newton search and continuation from non-{\eqb} states
that approximately balanced the mechanisms highlighted by the self-sustained process.
As noted in \cite{W03}, these solutions, and \cite{CB92}'s {\eqba} of plane
Couette flow with Rayleigh-Benard convection, are homotopic to the Nagata
{\eqba} under smooth transformations in the flow conditions. 
\cite{FE03} and \cite{WK04} carried the idea of a self-sustaining process
over to pipe flow and applied Waleffe's continuation strategy to derive
families of traveling-wave solutions for pipes. Traveling waves for \pCf\ were
computed by \cite{N97} using a continuation method.  Later, traveling waves for pressure-driven
channel flow were obtained by \cite{IT01} with a shooting method. The first
short-period unstable periodic solution of Navier-Stokes were computed by
\cite{KawKida01}. Recently, \cite{Visw07b} has computed {\em relative}
periodic orbits (orbits which repeat themselves with a translation) and further \po s of
\pCf\ that exhibit break-up and reformation of roll-streak structures.

The exact solutions described above turn out to be remarkably similar
in appearance to coherent structures observed in DNS and experiment.
\cite{W01} coined the term `exact coherent structures' to emphasize
this connection. The {\ubranch} solution, for example, captures
many statistical features of turbulent \pCf\ and appears remarkably
similar to the roll-streak structures observed in direct numerical
simulations (compare \reffig{f:uHKWtypical}\textit{(b)} to
\reffig{f:LBNBUBfields}\textit{(c)}).
\cite{W03} showed that the upper and lower-branch {\eqba} appear at
lowest Reynolds number with streak spacing of $100^+$ wall units, an
excellent match to that observed in \cite{KRSR67}. The periodic orbits of
\cite{KawKida01} and \cite{Visw07b} appear to be embedded in \pCf's
natural ergodic measure, and most of them capture basic statistics
more closely than the {\eqba}. In pipe flow, high speed streaks that
match the traveling-wave solutions in cross-section have been observed
in beautiful experiments using stereoscopic particle image velocimetry
(\cite{science04,Busse04,PhysWorld04}). Additionally,
there is preliminary evidence that the instabilities of these exact
solutions play important {\em dynamic} roles. The relevance of steady
solutions to sustained turbulence and transition to turbulence is discussed
in \cite{W03} and \cite{JKSNS05}. The stable manifold of the %Nagata-Waleffe
\lbranch\ solution is conjectured to control the basin boundary between the
turbulent and laminar attractors (\cite{WW05,WGW07,Visw07a}). \cite{Kerswell07}'s
numerical simulations have established that lower-branch traveling waves as
act as similar boundaries in pipe flow, and that turbulent fields make
occasional visits to the neighborhoods of traveling waves.
                
Together, these results form a new way of thinking about coherent structures
and turbulence: (a) that coherent structures are the physical images of the
flow's least unstable invariant solutions, (b) that turbulent dynamics consists
as a series of transitions between these states, and (c) that intrinsic
low-dimensionality in turbulence results from the low number of unstable modes
for each state (\cite{W02}). The long-term goals of this research program are to
develop this vision into quantitative, predictive description of moderate-{\Reynolds}
turbulence, and to use this description to control flows and explain their statistics.
Much of this has already been accomplished in the simpler context of the
Kuramoto-Sivashinsky equation (\cite{Christiansen:97,SCD07}).

In this paper, we take a few steps towards realizing these goals in the case of
plane Couette flow. In \refsect{s:review} we review the physical characteristics and
symmetries of plane Couette flow. \refSect{s:equilibria} discusses the computation of
invariant solutions and their eigenvalues and presents (a) a new {\eqb} solution
of plane Couette and (b) the linear stability analysis of this and the lower and
upper-branch {\eqba}. These computations set the stage for the main advance
reported in this paper, visualization and exploration of the \statesp\ of
moderate-\Reynolds\ \pCf, undertaken in \refsect{s:geometry}. The combination of
\eqb\ solutions, linear stability analysis, and {\stateDsp} portraiture reveals
previously unseen dynamical connections amongst the known invariant solutions of
\pCf. Particularly beautiful and unexpected are the discrete symmetry enforced
interrelations between unstable manifolds manifest in
\reffig{f:LBNBUB_portrait_a}-\reffig{f:turbulence}.

%% file: PCFreview.tex
% PCFreview.tex
% $Author: gibson $ $Date: 2008-02-11 00:38:43 +0530 (Mon, 11 Feb 2008) $

% \section{\PCf}
% \label{s:review}

\PCf\ is comprised of an incompressible viscous fluid confined
between two infinite parallel plates moving in-plane at constant
velocities. We take the length scale $L$ to be half the distance
between the walls and the velocity scale $U$ to be half the
relative wall velocity. After nondimensionalization and
absorption of fluid density into the pressure field, the
Navier-Stokes equations take the form
\[
    \frac{\partial {\bu}}{\partial t}
    + {\bu}\cdot{ \bnabla\bu}
=
    - { \bnabla} p
    + \frac{1}{\Reynolds}
        { \bnabla}^2 {\bu} \,, \quad \nabla \cdot \bu = 0
\,,
\]
where the Reynolds number is defined as $\Reynolds = UL/\nu$ and $\nu$ is
the kinematic viscosity of the fluid.
The plates move at speed $\pm 1$ along the `streamwise' $x$-axis,
the direction normal to the plates is the `wall-normal' $y$-axis,
$y \in [-1,+1]$, and the in-plane $z$-axis, normal to the plate
velocity, is referred to as `spanwise.'  The $x,y,z$ unit vectors are
$\ex, \ey, \ez$. (We use boldface to indicate vectors
in three spatial dimensions.)  The velocity field $\bu$ has
streamwise, wall-normal, and spanwise components $\bu = [u,v,w]$; the
velocity at point $\bx$ and time $t$ is
$\bu(\bx, t) = [u,v,w](x,y,z,t)$.
The no-slip boundary conditions at the walls are
$\bu(x,\pm 1, z)= [0,\pm 1,0]$.
Numerical computations replace the infinite $x$ and $z$ directions
with a periodic cell of lengths $L_x$ and $L_z$, or equivalently,
the fundamental wave\-numbers $\alpha = 2 \upi /L_x$ and
$\gamma= 2 \upi /L_z$.
We denote the periodic domain of the cell by
$\Omega = [0, L_x] \times [-1, 1] \times [0, L_z]$ or simply
$\Omega = [L_x, 2, L_z]$.
We assume that the spatial mean of the pressure gradient is
zero, \ie there is no pressure drop across the cell in $x$ or
$z$.

Replacing $\bu$ with $\bu + y \, \ex$ recasts Navier-Stokes
in terms of the difference of the velocity from laminar flow:
\beq
    \frac{\partial { \bu}}{\partial t}
    + y  \frac{\partial { \bu}}{\partial x}
    + v \, \ex
    + { \bu}\cdot{ \bnabla \bu}
=
    - { \bnabla} p
    + \frac{1}{\Reynolds}
        { \bnabla}^2 { \bu } \,, \quad \nabla \cdot \bu = 0
\,.
\ee{NavStokesDev}
The difference $\bu$ satisfies Dirichlet conditions at the walls,
 $\bu(x, \pm 1, z) = 0$. Henceforth we refer to the difference $\bu$ as
`velocity' and $\bu + y \, \ex$ as `total velocity,' and we take
\refeq{NavStokesDev} as the Navier-Stokes equations for \pCf.

%%%%%%%%%%%%%%%%%%%%%%%%%%%%%%%%%%%%%%%%%%%%%%%%%%%%%%%%%%
\begin{figure}
\centering
{\small (a)}\includegraphics[height=2in]{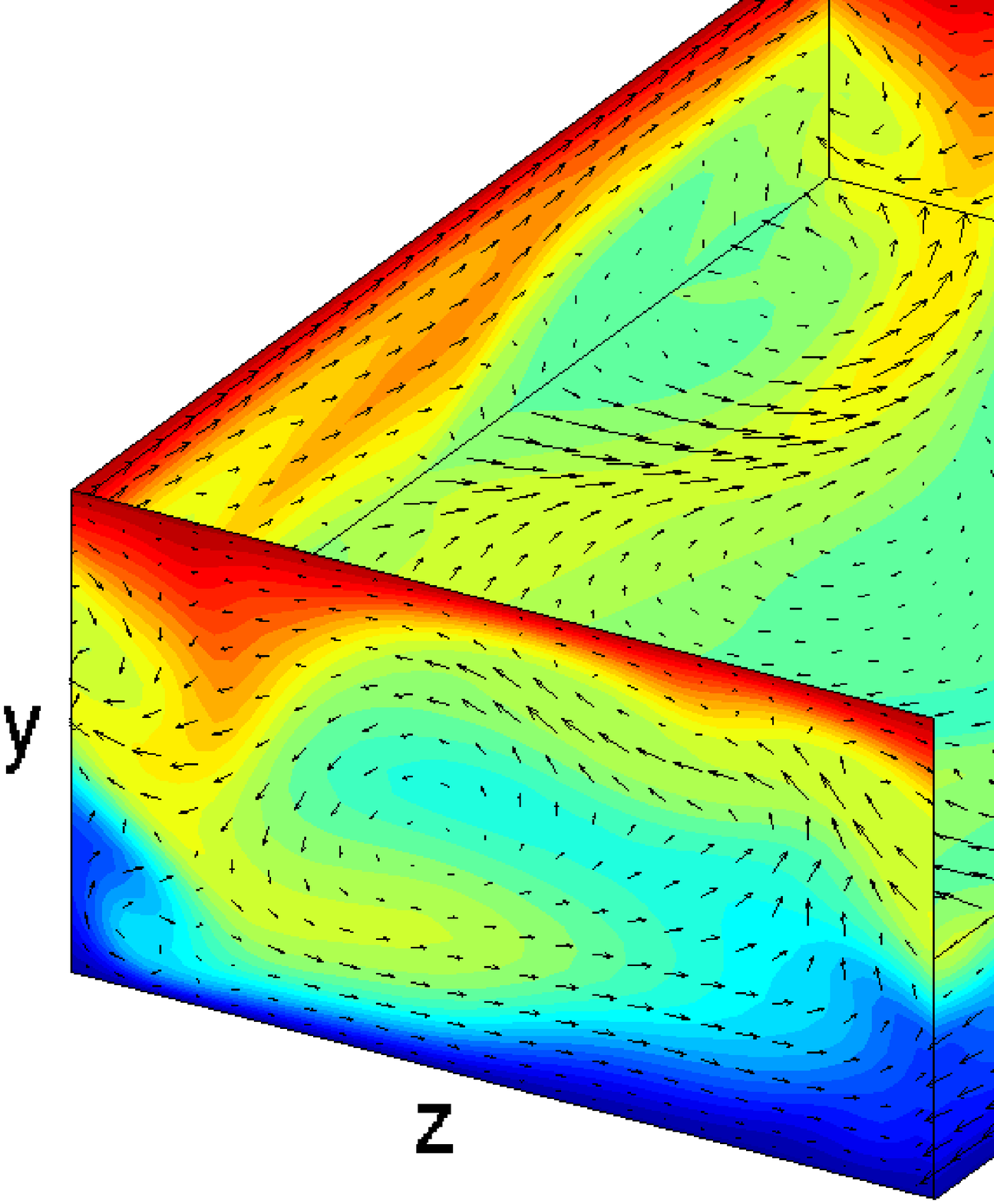}
{\small (b)}\includegraphics[height=2in]{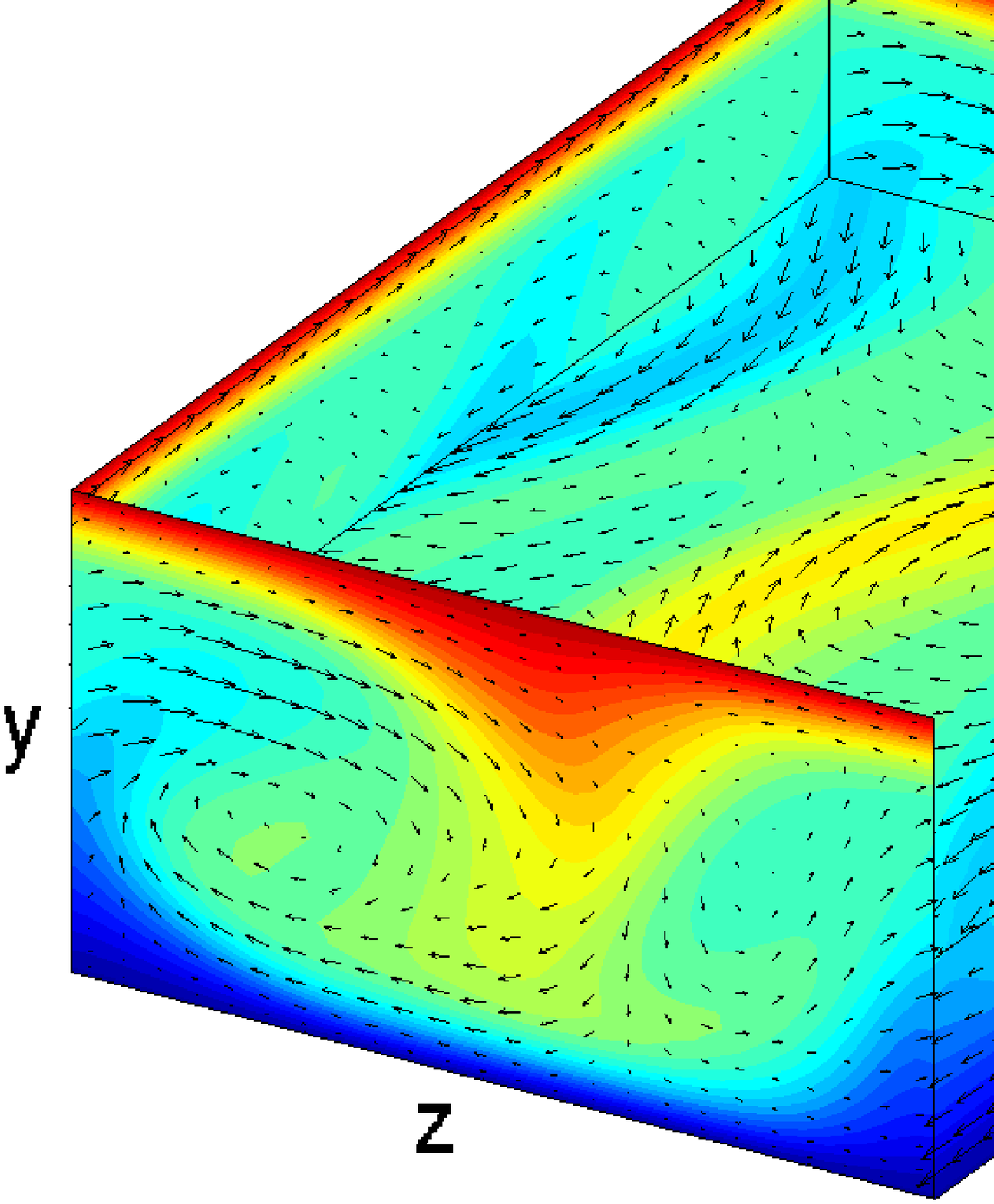}
\caption{
{\bf Snapshots of plane Couette turbulence at
$\Reynolds = 400$.}
Velocity fields $(u,v,w)$ in a periodic cell of size $\Omega =
[7\upi/4,2,6\upi/5]$ (\cite{HaKiWa95}) are shown with arrows for
in-plane velocity and \colorcomm{a colormap}{grayscale} for the streamwise velocity component
$u$: \colorcomm{red/blue}{light/dark} indicates $u=\pm1$; \colorcomm{green}{white}, $u=0$. 
The upper wall at $y=1$
and the upper half of the fluid is cut away to show the velocity in the
$y=0$ midplane. The two snapshots shown are different instants from
a simulation initiated with a random pertubation, selected to show
(a) minimum and (b) maximum organization in the turbulent field.
In particular, (b) resembles the upper-branch {\eqb} shown
in \reffig{f:LBNBUBfields}\textit{(c)}.}
\label{f:uHKWtypical}
\end{figure}
%%%%%%%%%%%%%%%%%%%%%%%%%%%%%%%%%%%%%%%%%%%%%%%%%%%%%%%%%%

Plane Couette flow is the simplest of all shear flows, and it is here
that roll-streak structures take their simplest form. For moderate values of $\Reynolds$,
the rolls span the full distance between the walls, whereas in
channel and boundary-layer flows such structures are bounded by a wall
on one side and open flow on the other. \refFig{f:uHKWtypical} shows two typical
velocity fields from a simulation in the `HKW' cell $\Omega = [7\upi/4,2,6\upi/5]$
at $\Reynolds = 400$.
The numerical simulations of \cite{HaKiWa95} indicate that this is roughly
the smallest cell and Reynolds number that sustains turbulence for long time
scales. Roll-streak structures are
evident, particularly in \reffig{f:uHKWtypical}\textit{(b)}. The rolls circulate
high-speed fluid towards the walls and low-speed flow away; the
resulting streaks of high-speed fluid near the walls dramatically increase
drag compared to laminar flow. For example, the power input needed to
maintain constant wall velocity in \pCf\ increases by a factor of three
if the flow goes turbulent (see \reffig{f:KKIvsD}\,(\textit{e})). The
practical importance of roll-streak dynamics derives from their role
in momentum transfer and turbulent energy production and their generic
occurrence in wall-bounded shear flows.

Except for \reffig{f:uHKWtypical} and parts of \reffig{f:KKIvsD}, the results
in this paper are for $\Reynolds = 400$ and $\Omega = [2\upi/1.14,2,4\upi/5]$,
first studied in \cite{W02}. This cell matches the HKW cell $[7\upi/4,2,6\upi/5]$
closely in $x$ ($7/4 \approx 2/1.14$). The $z$ length scale $L_z = 4\upi/5$ was
chosen as a compromise between $L_z = 6 \upi /5$ of the HKW cell (which sustains
turbulence for long time scales but has {\eqba} only with doubled period in $z$)
and its first harmonic $L_z = 3 \upi /5$ (which has {\eqba} at the fundamental
harmonic, but tends to decay to laminar flow). Simulations for these parameters tend
to decay to the laminar state in within several hundred nondimensionalized time units
$L/U$, but the transient dynamics serves well to illustrate our invariant manifolds
construction. Whether a given cell size sustains turbulence indefinitely is a subtle
dynamical issue: \cite{SE97} and \cite{Schmi99} observe only chaotic transients in their studies.

\subsection{Energy transfer rates}

\begin{figure}
\centerline{
\includegraphics[width=0.31\textwidth]{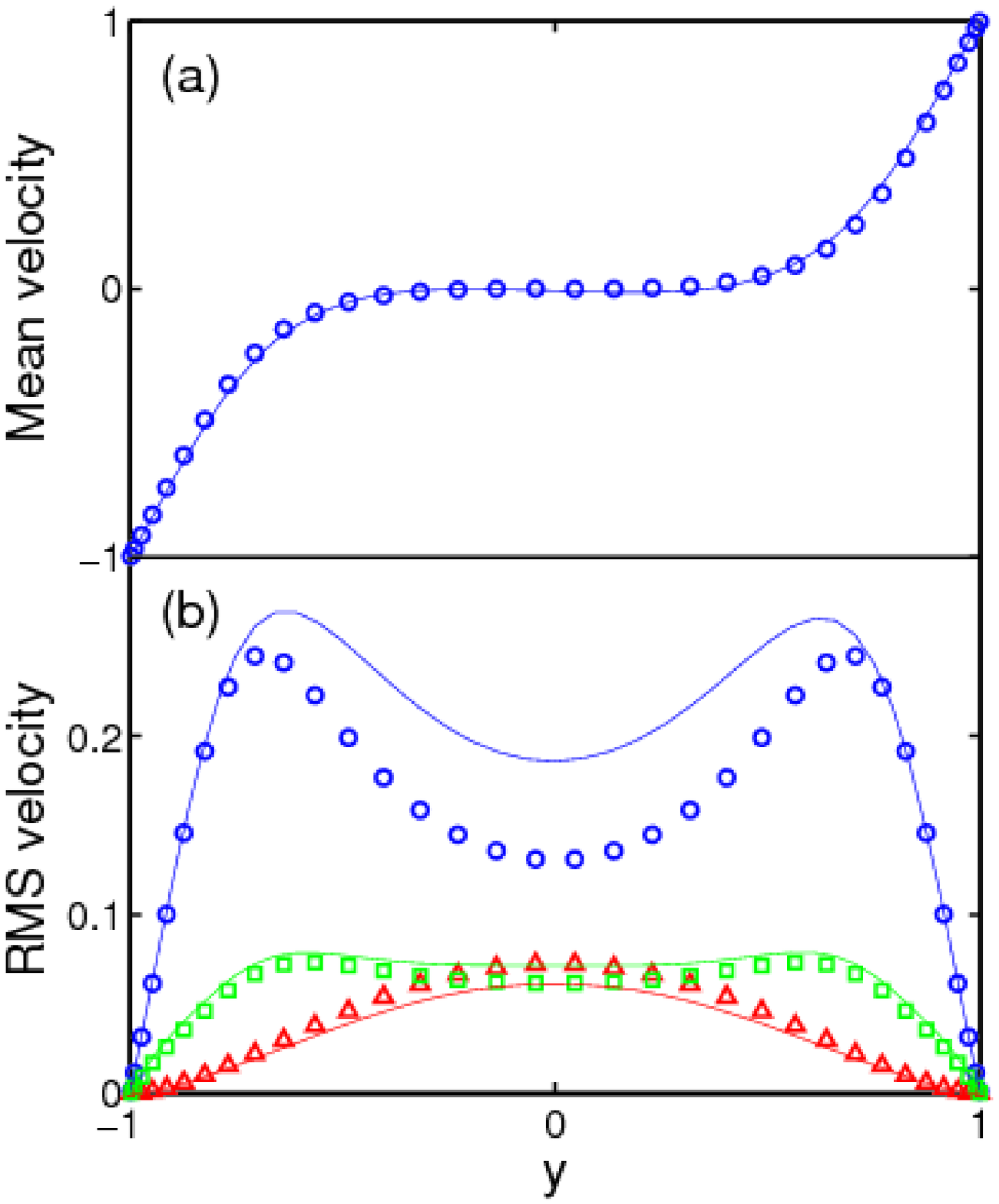}
%\ifpdf
%\includegraphics[angle=90,width=0.31\textwidth]{kk_pcf_u_profiles-traced.eps}
%\else
  \includegraphics[width=0.31\textwidth]{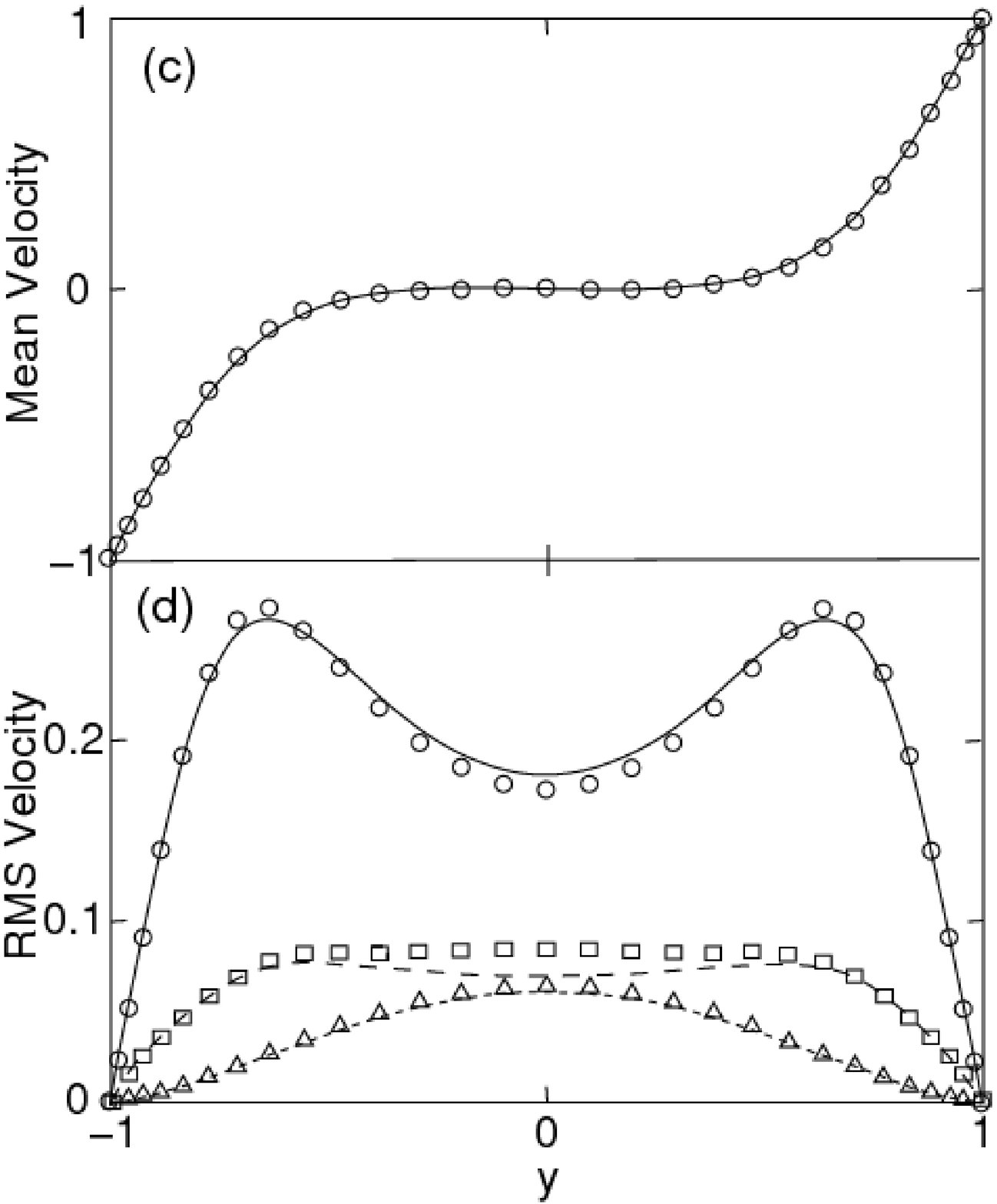}
%\fi
\includegraphics[width=0.32\textwidth]{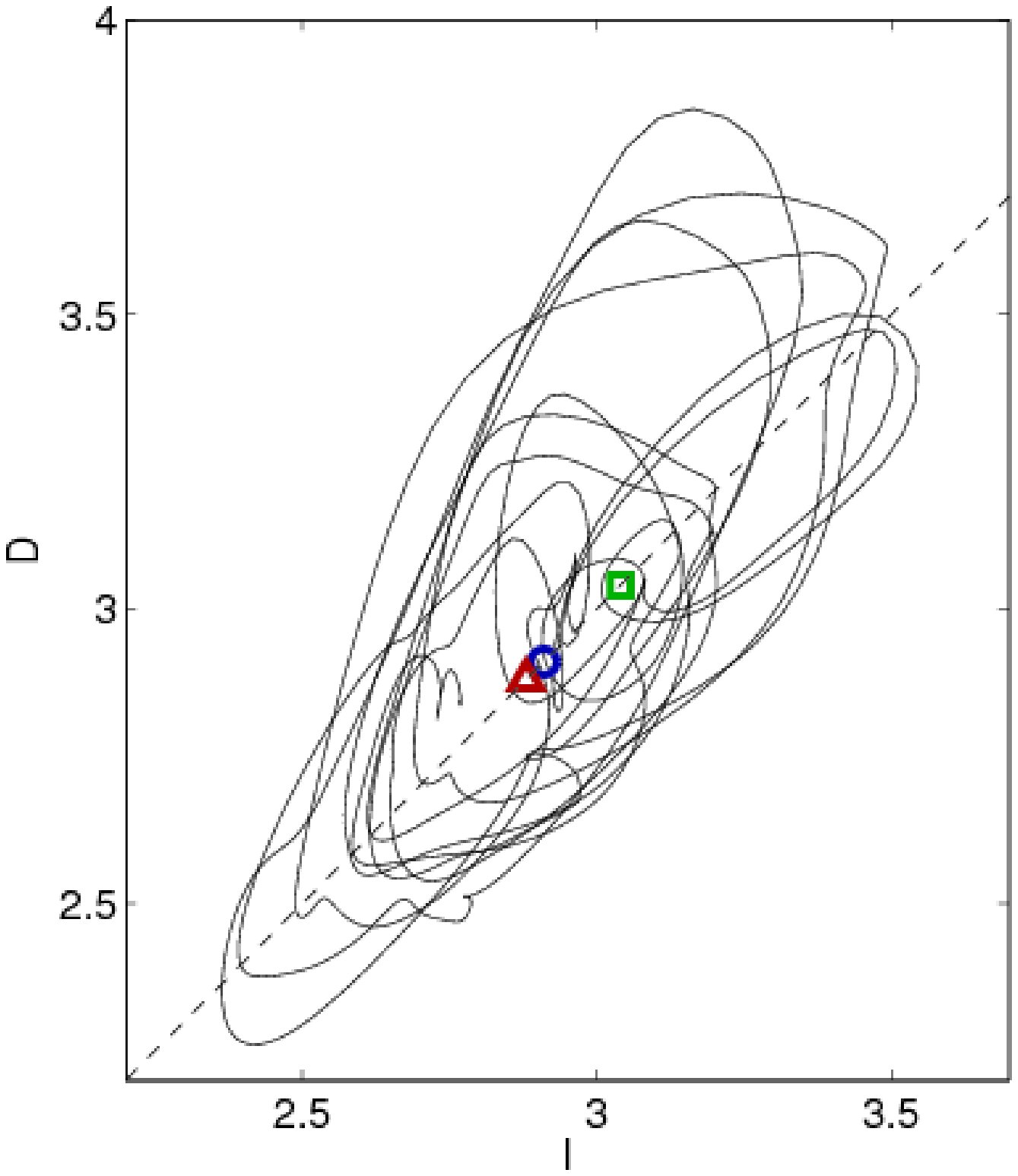}
}
\caption{
(a,b) Spatial-mean and spatial-RMS velocity profiles of the 
\cite{N90} upper-branch \eqb\ 
(symbols) in $\Omega = [2\upi/1.14,2,4\upi/5]$ against temporal mean and 
RMS profiles for sustained turbulent DNS data in $[7\upi/4,2,6\upi/5]$ 
(lines). $\Reynolds =400$ for both.
(c,d) Temporal mean and RMS velocity profiles for the \cite{KawKida01} 
$\Reynolds = 400$ periodic solution (symbols), against the same turbulent 
data shown in (a,c). 
In (a,c), mean values of $u$ are marked with $\circ$; in (b,d) RMS values
of $u,v,w$ are marked with $\circ, \triangle, \square$.
(e) $I$ versus $D$ for a typical trajectory of sustained
turbulence in the $[7\upi/4,2,6\upi/5]$ cell (lines),
mean dissipation rate over \cite{KawKida01} periodic orbit 
$\timeAver{D} = 2.91$ ($\circ$), mean turbulent dissipation 
$\timeAver{D} = 2.89$ ($\triangle$), and upper-branch \eqb\ 
$D_{\text{\tiny{UB}}} = 2.89$ ($\square$).
The laminar \eqb\  dissipation rate is $D=1$ (not shown).
}
\label{f:KKIvsD}
\end{figure}

The kinetic energy density $E$, the bulk dissipation rate $D$, and the power input $I$
of total velocity field of plane Couette flow are given by
\begin{align}
E(t) &=  \frac{1}{V}  \int_\Omega   \!d\bx\,  \frac{1}{2} |\bu + y \, \ex|^2 \label{KinEnerg} \\
D(t) &= \frac{1}{V} \int_\Omega \!d\bx\, |\bnabla \times (\bu + y \, \ex)|^2 \label{BulkRate} \\
I(t) &= 1+ \frac{1}{2A} \int_A \!  dx\, dz\,
    \left( \frac{\partial u}{\partial y} \Bigl\lvert_{y=1}  + \frac{\partial u}{\partial y} \Bigl\lvert_{y=-1}
    \right)
\,,
\label{InDrag}
\end{align}
where $V = 2 L_x L_z$ and $A =  L_x L_z$. The normalizations are chosen so that
$D = I = 1$ for laminar flow and $\dot{E} = I - D$.
%Energy is injected through the
%motion of the walls and dissipated by viscosity, so that the power input $I$ is the
%power needed to maintain constant wall velocity $U=\pm 1$.
\refFig{f:KKIvsD}\textit{(e)} shows $I$ {\em vs.} $D$ for a turbulent trajectory and
several invariant solutions. \Eqba\ and \reqba\ must fall on the line $I=D$
where these two quantities are in balance. The energy input and dissipation rates
must also balance in averages over a single period of a \po\ or \rpo\ $p$,
$D_p = 1/\period{p} \int_0^\period{p} \!\! dt \, D(t)
   = 1/\period{p} \int_0^\period{p} \!\! dt \, I(t) = I_p$,
as well as for long-term averages, $\timeAver{I(t)}=\timeAver{D(t)}$.

%%%%%%%%%%%%%%%%%%%%%%%%%%%%%%%%%%%%%%%%%%%%%%%%%%%%%%%%%%%%%
%%%%%%%%%%%%%%%%%%%%%%%%%%%%%%%%%%%%%%%%%%%%%%%%%%%%%%%%%%%%%

We note that the turbulent trajectory in \reffig{f:KKIvsD} stays clear of
the much lower dissipation rates 
$D_{\text{\tiny LB}} = 1.429$ and 
$D_{\text{\tiny NB}} = 1.454$
of the {\eqb} solutions \uLB\ and \uNB\
(see \refsect{s:equilibria_subsection}),
so these {\eqba} are far from the turbulent attractor.
The energy, the dissipation rate, and mean and RMS velocity profiles
of the \cite{N90} upper-branch \eqb\ and 
the \cite{KawKida01} periodic orbit are all
numerically close to the long-term turbulent averages.
This {\em suggests} that the solutions play an important role
in turbulent dynamics, but turbulent statistics do not simply follow
from the properties of one or two solutions. On the contrary,
periodic orbit theory shows that the statistics of
dynamical systems are given by sums over {\em hierarchies} of
periodic orbits, with weights determined by the orbits' lengths
and stabilities (\cite{DasBuch}).

%% file: PCFsymm.tex
% PCFsymm.tex
% $Author: gibson $ $Date: 2008-02-10 12:42:47 +0530 (Sun, 10 Feb 2008) $

\PCf\ is invariant under two discrete symmetries $\sigma_1,\sigma_2$ and a
continuous two-parameter group of translations $\tau(\shift_x, \shift_z)$:
\begin{align}
\sigma_1 \, [u,v,w](x,y,z) &= [u, v,-w](x,y,-z) \nnu \\
\sigma_2 \, [u,v,w](x,y,z) &= [-u,-v,w](-x,-y,z)  \label{reflSfit}\\
\tau(\shift_x, \shift_z)[u,v,w](x,y,z) &= [u,v,w](x+\shift_x,y,z+\shift_z) \nnu\,.
\end{align}
The Navier-Stokes equations and boundary conditions are invariant under any
symmetry $s$ in the group generated by these symmetries:
$\partial (s \bu) / \partial t = s (\partial \bu / \partial t)$.

The \cite{N90} lower and upper-branch \eqba\ \uLB\ and \uUB\ 
are invariant under action of  the subgroup $S = \{1, s_1, s_2, s_3\}$,
where $s_1 = \tau(L_x/2,0) \, \sigma_1$, $s_2 = \tau(L_x/2,L_z/2) \, \sigma_2$,
and $s_3 = s_1 s_2$. That is, $s \, \uLB = \uLB$ and $s \, \uUB = \uUB$ for $s \in S$.
The $s_1$ and $s_2$ symmetries are referred to as the `shift-reflect' and
`shift-rotate' symmetries.
%\JFG{Track down first use, cite. PC: forget it}
The group actions on
velocity fields $\bu$ are given by
\begin{align}
s_1 \, [u, v, w](x,y,z) &= [u, v, -w](x+L_x/2,\, y,\, -z) \nnu \\
s_2 \, [u, v, w](x,y,z) &= [-u, -v, w](-x+L_x/2,\,-y,\,z+L_z/2) \label{shiftRot} \\
s_3 \, [u, v, w](x,y,z) &= [-u,-v,-w](-x,\, -y,\, -z+L_z/2) \nnu \,.
\end{align}

We denote the space of velocity fields that satisfy the kinematic
conditions of \pCf\ by
\begin{align}
 \bbU  &= \{\bu \; | \; \grad \cdot \bu = 0,
               \; \bu(x, \pm 1, z) = 0, \; \bu(x, y, z) = \bu(x+L_x, y, z) = \bu(x, y, z + L_z)\}  \,.
\label{Udefn}
\end{align}
and the $S$-invariant subspace (\cite{golubitsky2002sp}) of $\bbU$ by
\begin{align}
\bbU_S  &= \{\bu \in \bbU  \: | \;
              s_j \bu = \bu\,, \;\;  s_j \in S \}
              % \bu = \frac{1}{4} (1 + s_1 + s_2 + s_3)\,\bu \}
\label{symmSubspU}
\end{align}
$\bbU_S$ is a flow-invariant subspace of $\bbU$ since $S$ symmetry is preserved
by evolution under the Navier-Stokes equations.

A second important subgroup is the group of half-cell translations
$T = \{1, \tau_x, \tau_z, \tau_{xz}\}$, where $\tau_x = \tau(L_x/2,0)$,
$\tau_z = \tau(0,L_z/2)$, and $\tau_{xz} = \tau_x \tau_z$. In general,
the continuous translation $\tau(\ell_x, \ell_z)$ maps each state $\bu$
into a 2-torus of dynamically equivalent states, and the group
$\{1, \sigma_1, \sigma_2, \sigma_1 \sigma_2\}$ maps these into four
dynamically equivalent 2-tori. For $\bu \in \bbU_S$, the four tori coincide,
and the torus intersects $\bbU_S$ at the four points $\tau \bu$, $\tau \in T$.
(Since elements of $T$ commute with those of $S$, $\bu \in \bbU_S$ implies
$\tau \bu \in \bbU_S$ for $\tau \in T$.) For example, the upper-branch
{\eqb} \uUB\ appears within $\bbU_S$ in four distinct half-cell translations,
namely $\uUB, \, \tau_x \uUB, \, \tau_z \uUB,$ and $\tau_{xz} \uUB$.

%% file: equilibria.tex
% PCgeom.tex
% $Author: gibson $ $Date: 2008-02-11 00:38:43 +0530 (Mon, 11 Feb 2008) $

%\subsection{Invariant solutions}

Let $\vField(\bu)$ represent the \NSe\ \refeq{NavStokesDev}
for $\bu \in \bbU$ \refeq{Udefn} and $\bff^t$ its time-$t$
forward map
\begin{align}
\pd{\bu}{t} = \vField(\bu)
    \,, \qquad
    \bff^t(\bu) = \bu + \int_0^t \! d \tau\,  \vField(\bu)
\,.
\label{symbolicNS}
\end{align}
$\vField(\bu)$ admits of invariant solutions of the following types:
\begin{align}
\vField(\uEQ) &= 0  &&  \text{\eqb\ or steady state } \uEQ
    \nnu \\
\vField(\uTW) &= - {\bf c} \cdot \grad \uTW &&
       \text{relative \eqb\ or traveling wave \uTW, velocity }{\bf c} \nnu \\
\bff^\period{p}(\bu_p) &= \bu_p && \text{\po\ $p$ of period } \period{p}
    \nnu \\
\bff^\period{p}(\bu_p) &=
        \tau_p \, \bu_p
             && \text{\rpo, period $\period{p}$, shift }
             \tau_p=\tau(\ell_x,\ell_z)
    \,.
\label{contInvSolns}
\end{align}
\Reqba\ and \rpo\ solutions are allowed due to the continuous
translation symmetry $\tau(\shift_x, \shift_z)$; for traveling waves,
boundary conditions require ${\bf c} \cdot \ey = 0$.
We expect to see many more \rpo s than \po s because a trajectory
that starts on and returns to a given torus is unlikely
to intersect it at the initial point, unless forced
to do so by a discrete symmetry.
This indeed is the case for other PDEs with continuous symmetries,
such as the complex Ginzburg-Landau equation (\cite{LBHM05})
and the \KSe\ (\cite{SCD07}).
Restriction to the $S$-invariant
subspace $\bbU_S$ defined in \refeq{symmSubspU}
eliminates \reqba\ and \rpo s.
In what follows we focus mostly on dynamics within $\bbU_S$.

\subsection{Finite representation}
\label{s:finite_rep}

Computing the exact solutions and stability modes of plane Couette
flow requires a finite but fully-resolved discretization of the constrained partial
differential and integral equations represented by \refeq{symbolicNS} and
\refeq{contInvSolns}. We investigated two approaches to discrete representation.
In the first approach
%, following previous literature (citations),
the vector $\ucoeff \in \mathbb{R}^d$ was
formed by breaking the complex spectral expansion coefficients of
a CFD algorithm into real and imaginary parts and then selecting from
these a set of linearly independent real-valued coefficients. Our CFD
algorithm, {\tt channelflow.org}, is based on the velocity-pressure
algorithm of \cite{Kleiser80}) with expansions
\beq
\bu(\bx,t) = \sum_{j=-J}^J \sum_{k=-K}^K \sum_{\ell=0}^L \sum_{m=1}^3
   \hat{u}_{jklm} \,T_{\ell}(y) \,
   e^{2\upi i (jx/L_x + kz/L_z)}\, \ex_m \,,
\ee{CFDexpansion}
where the $T_{\ell}$ are Chebyshev polynomials and $(\ex_{1},\ex_{2},
\ex_{3}) = (\ex,\ey,\ez)$ unit vectors. The algorithm employs a Chebyshev
tau method and tau correction for enforcement of incompressibility and
boundary conditions, third-order semi-implicit backwards-differentiation
time-stepping, dealiasing in the $x,z$ transforms, and a variety of
methods for calculating the nonlinear term. The expansion
\refeq{CFDexpansion} retains a number of linearly dependent terms, due
to complex symmetries and the run-time enforcement of the
incompressibility and boundary conditions. Intimate knowledge of the
CFD algorithm and careful accounting is required to determine
the precise value of the dimension of the linearly independent set
and a self-consistent method of converting back and forth between the
\stateDsp\ vector $u$ and the expansion coefficients of $\bu$. For
our CFD algorithm and an $N_x \times N_y \times N_z$ grid, $d$ is slightly
less than $2 N_x (N_y-2) N_z$. The accounting for velocity-vorticity
algorithms is somewhat simpler since incompressibility is eliminated at
the outset. For further technical details, please refer to \cite{Visw07b},
\cite{HalcrowThesis}, and documentation at {\tt channelflow.org}, or
contact the authors.

In the second approach, we explicitly constructed a set of
orthonormal, divergence-free, no-slip basis functions $\bPhi_n(\bx)$
and formed the \stateDsp\ vector $\ucoeff$ from the coefficients
$\hat{u}_n$ of the expansion $\bu(\bx) = \sum_{n=1}^d \hat{u}_n
\bPhi_n(\bx)$. This approach produces a mathematically simpler representation,
in that (1) all constraints are subsumed into the basis and eliminated
from further consideration, (2) an explicit second-order ODE of form
$\dot{\ucoeff} = F(\ucoeff)$ can be derived through Gal\"erkin projection of the
Navier-Stokes equation onto the basis set, and (3) with proper
normalization of the basis functions, the $L^2$ norm of the {\stateDsp}
vector $\ucoeff \in \mathbb{R}^d$ is the same as the $L^2$ energy norm
of the velocity field $\bu$. The downside is that the formulation of the
basis set is complicated, and it requires extra computation for
orthogonalization and transforms between \stateDsp\ vectors and the CFD
representation. We found no practical advantages to the orthonormal basis.
The results reported here were computed using CFD expansion coefficients
for the \stateDsp\ vector $\ucoeff$.

The choice of discretization $\ucoeff \in \mathbb{R}^d$ and CFD algorithm
implicity defines a $d$-dimensional dynamical system $\dot{\ucoeff} = F(\ucoeff)$.
\cite{Visw07b} showed invariant solutions and linear stability of $F$
can be computed efficiently with Krylov subspace methods and numerical
evaluation of the finite-time map $f^T : \ucoeff(t) \rightarrow
\ucoeff(t+T)$ with the CFD algorithm. Equilibria may computed as solutions
of $f^t(u) - u = 0$ for fixed $t$; and periodic orbits as solutions of
the same equation with varying $t$. Viswanath's algorithm for computing these
solutions involves a novel combination of Newton descent, GMRES solution of
the Newton equations, and `trust-region' limitation to the magnitude of
the Newton steps. The results reported in this paper, however, used straight
Newton-GMRES search, with no trust region modification. We will often discuss
\eqba\ and linear stability in terms of the flow $F$, with the understanding
that the computations are performed using the finite-time map $f^T$.

\subsection{Equilibria}
\label{s:equilibria_subsection}

%%%%%%%%%%%%%%%%%%%%%%%%%%%%%%%%%%%%%%%%%%%%%%%%%%%%%%%%%%%%%%%%%%
\begin{figure}
%\vspace*{-5pt}
\centerline{
{\small (a)} \hspace{-2mm}
\includegraphics[width=0.29\textwidth]{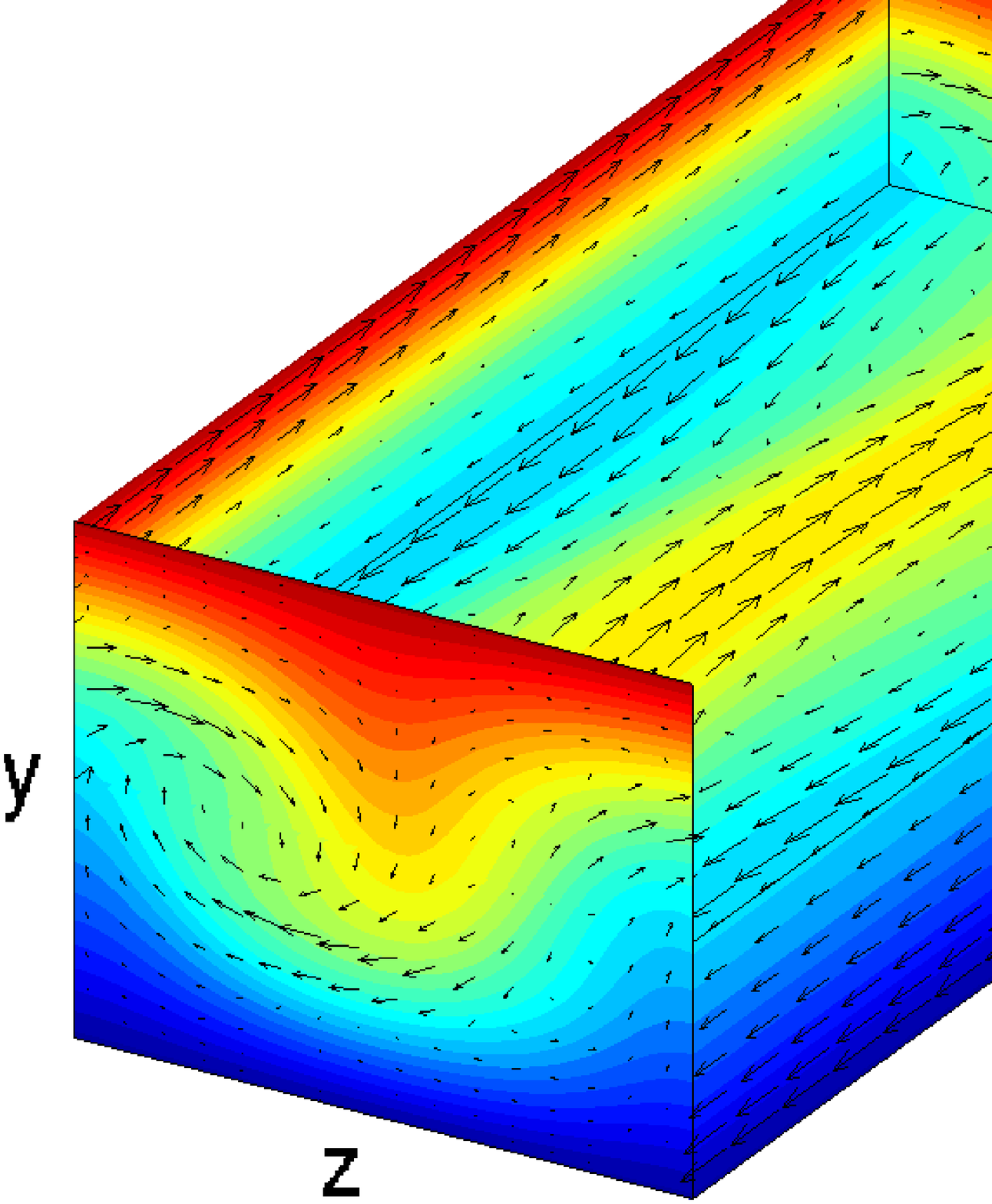}
{\small (b)} \hspace{-2mm}
\includegraphics[width=0.29\textwidth]{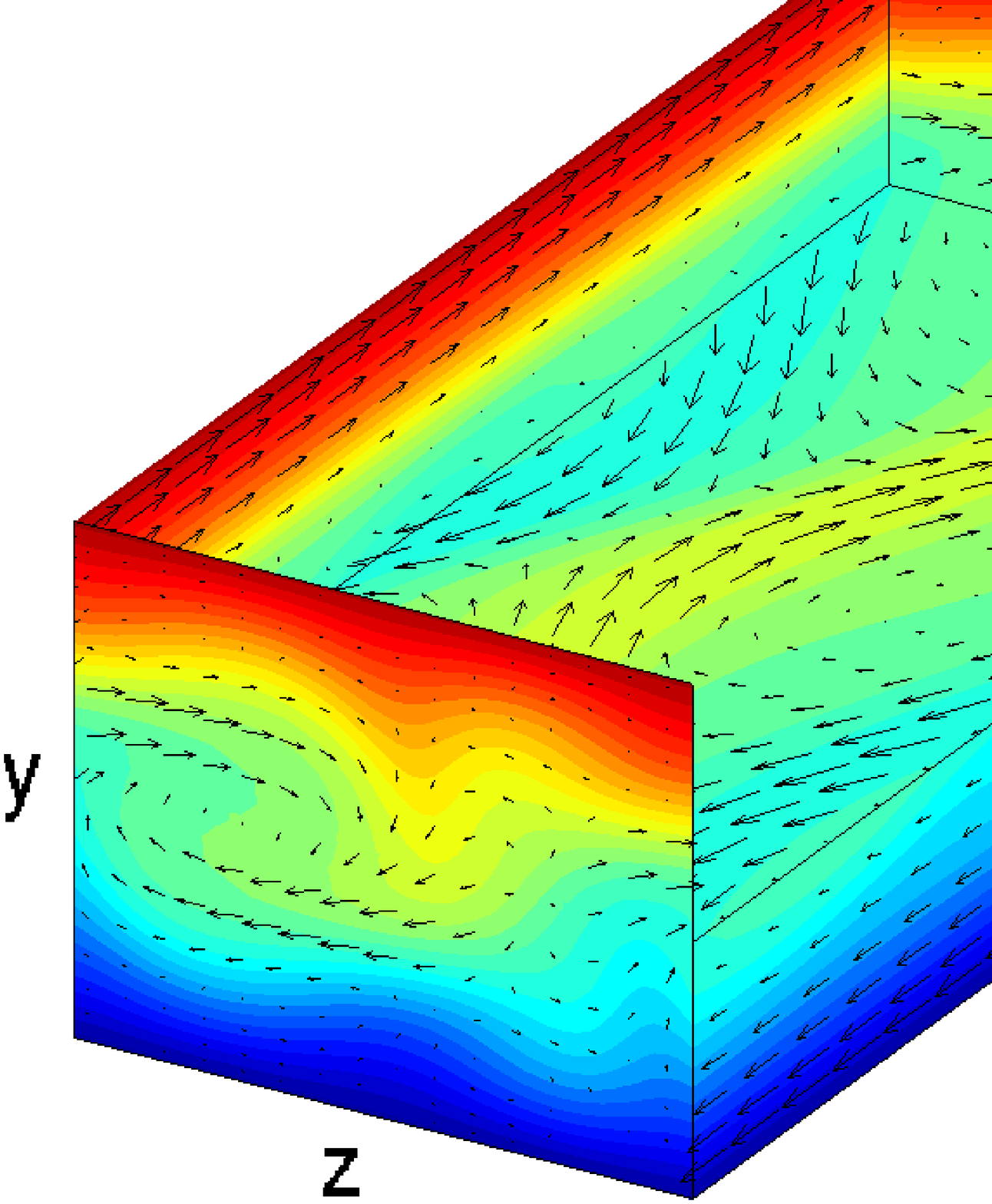}
{\small (c)} \hspace{-2mm}
\includegraphics[width=0.29\textwidth]{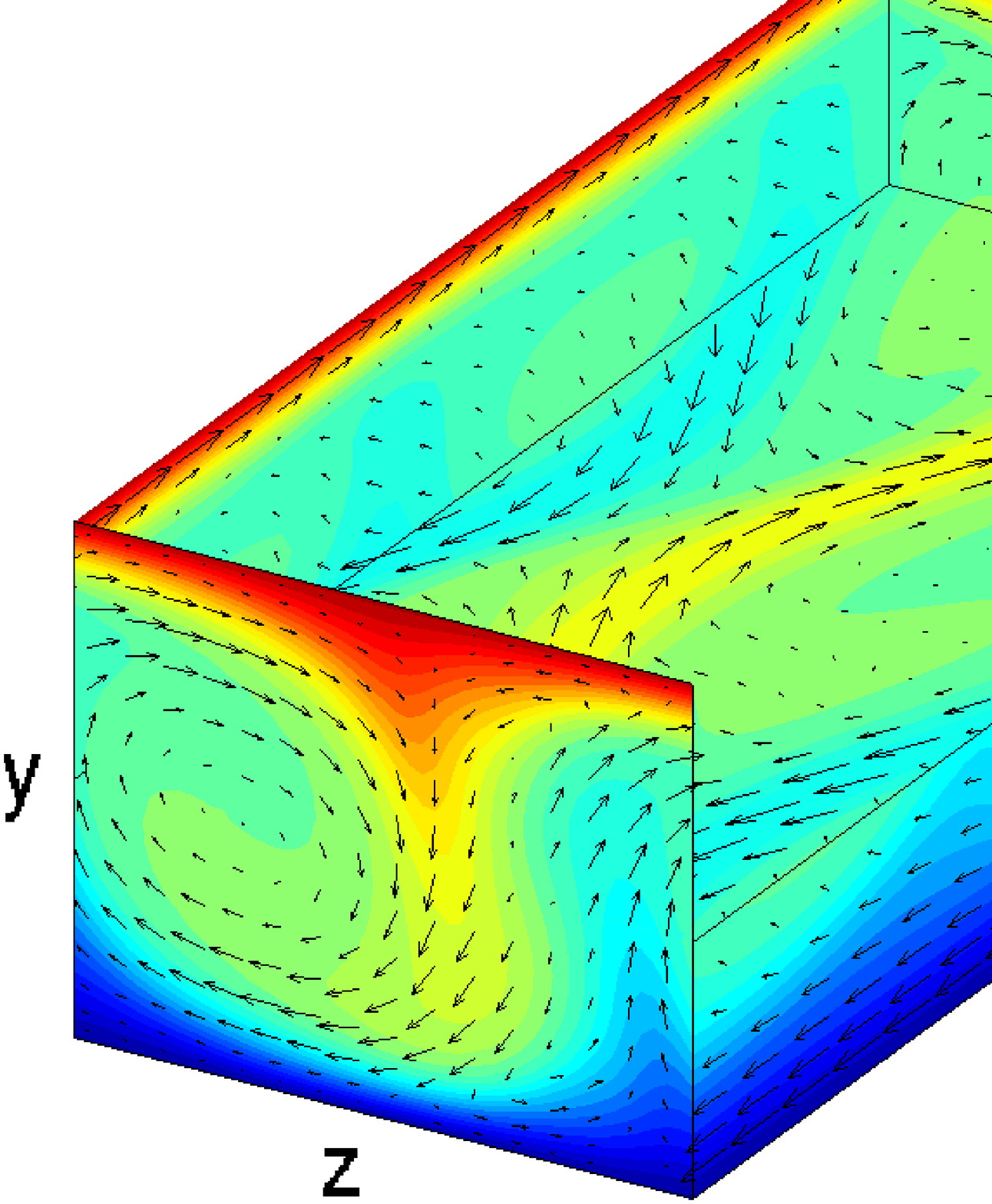}
}
%\vspace*{-5pt}
\caption{
{\bf Equilibrium solutions of plane Couette flow:}
(a) \uLB, the lower-branch {\eqb},
(b) \uNB, the `newbie' {\eqb}, and
(c) \uUB, the upper-branch {\eqb}.
$[L_x,L_y,L_z] = [2\upi/1.14,2,4\upi/5]$ and $\Reynolds = 400$.
The plotting conventions are the same as in \reffig{f:uHKWtypical}.
}
\label{f:LBNBUBfields}
\end{figure}
%%%%%%%%%%%%%%%%%%%%%%%%%%%%%%%%%%%%%%%%%%%%%%%%%%%%%%%%%%%%%%%%%%

The starting points of our exploration of plane Couette {\statesp}
are the \cite{N90} and \cite{W03} \uUB\ and \uLB\ \eqb\ solutions of
\pCf\ for $\Reynolds = 400$ and $[L_x,L_y,L_z] = [2\upi/1.14,2,4\upi/5]$,
provided in numerical form by Waleffe.
These fields employed an elliptical truncation of spectral coefficients
(requiring $j^2/J^2 + k^2/K^2 + l^2/L^2 \leq 1$ for expansions of the form
\refeq{CFDexpansion}) on a $32 \times 34 \times 32$ grid. We use Viswanath's
Newton-GMRES algorithm to increase the resolution to a rectangular truncation
($|j| \leq J, |k| \leq K, l \leq L$) on a $32 \times 35 \times 32$ grid. The
dimensionality of this finite representation is $d = 61\,506$. These recomputed
solutions satisfy \refeq{contInvSolns}
\begin{align}
F(\ucoeff) = 0
    \,, \qquad
    \ucoeff = f^t(\ucoeff)
    \,
\label{statespEqbEqn}
\end{align}
in discrete form. At this spectral resolution, the Newton-GMRES search can
reduce the residual of the discrete {\eqb} equations to $10^{-14}$,
but the truncated coefficients are of the order $10^{-6}$, and the residual
increases to the same level when the given solution is integrated at
higher resolution. The accuracy of the {\eqb} solutions is thus roughly
single-precision.

\refFig{f:LBNBUBfields} shows the \uLB\ and \uUB\ {\eqba} as velocity fields,
along with a third \eqb, \uNB, that was discovered in the course of
this investigation (\cite{HGC08}). We refer to the new \eqb\ as \uNB, pronounced
`newbie,' in keeping with the nomenclature for other \eqba. This
{\eqb} was found by initiating Newton-GMRES searches for zeroes
of the {\eqb} equations from points within the unstable manifolds
of \uLB\ and \uUB (see \refsect{s:geometry}).
A portion of the one-dimensional unstable manifold of \uLB,
shown in \reffig{f:LBNBUB_portrait_a}, appears to be strongly influenced
by a complex unstable eigenvalue of \uNB. Initial guesses along this
portion of the \uLB\ unstable manifold converge rapidly to \uNB, to 
several digits of accuracy in a few Newton steps.

\subsection{Linear stability of \eqba}
\label{s:linear_stability}

\begin{figure}
%\vspace*{-5pt}
\centerline{
{\small (a)}\!\! \includegraphics[width=0.44\textwidth]{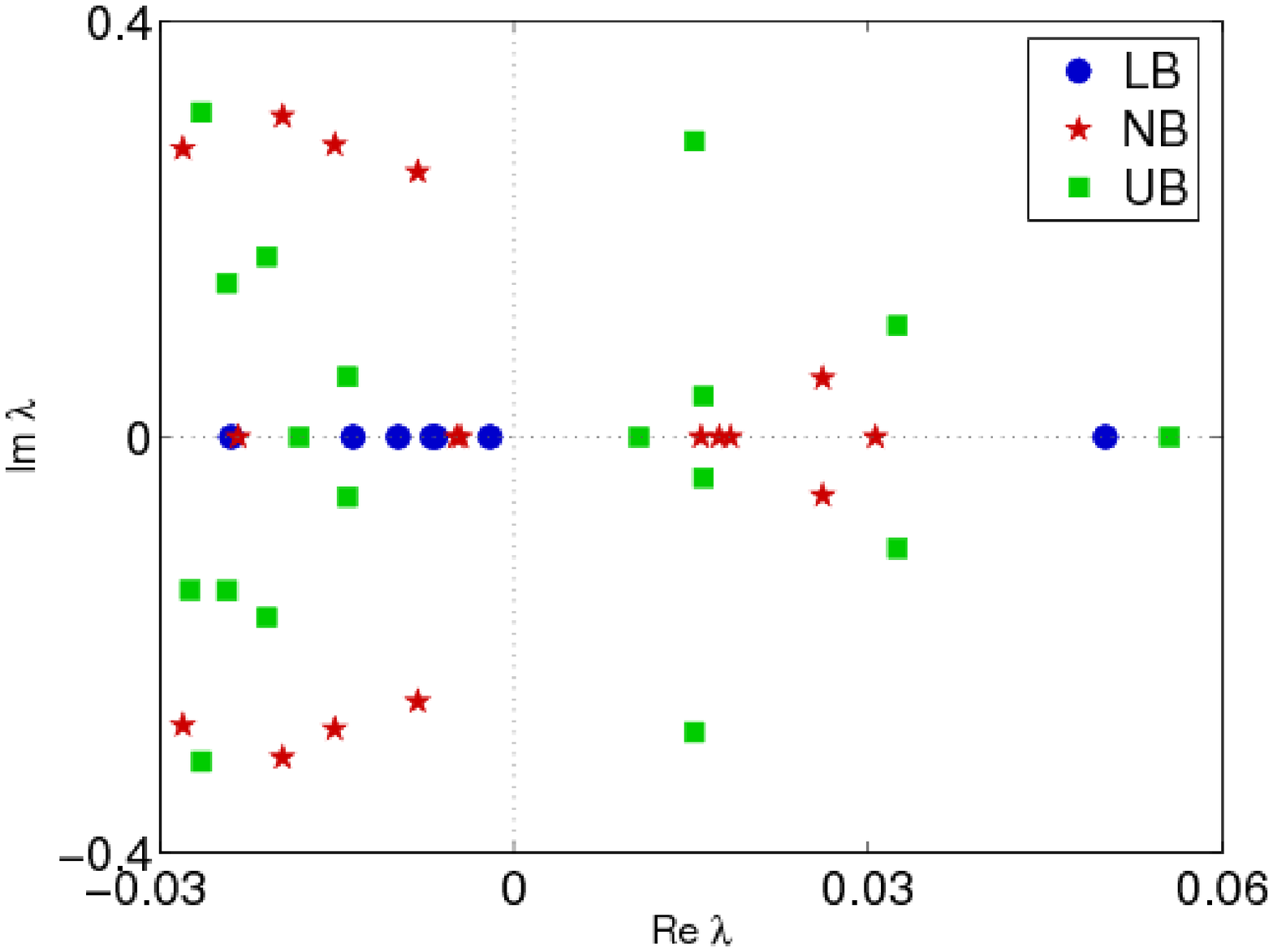}%
~~{\small (b)}\!\! \includegraphics[width=0.44\textwidth]{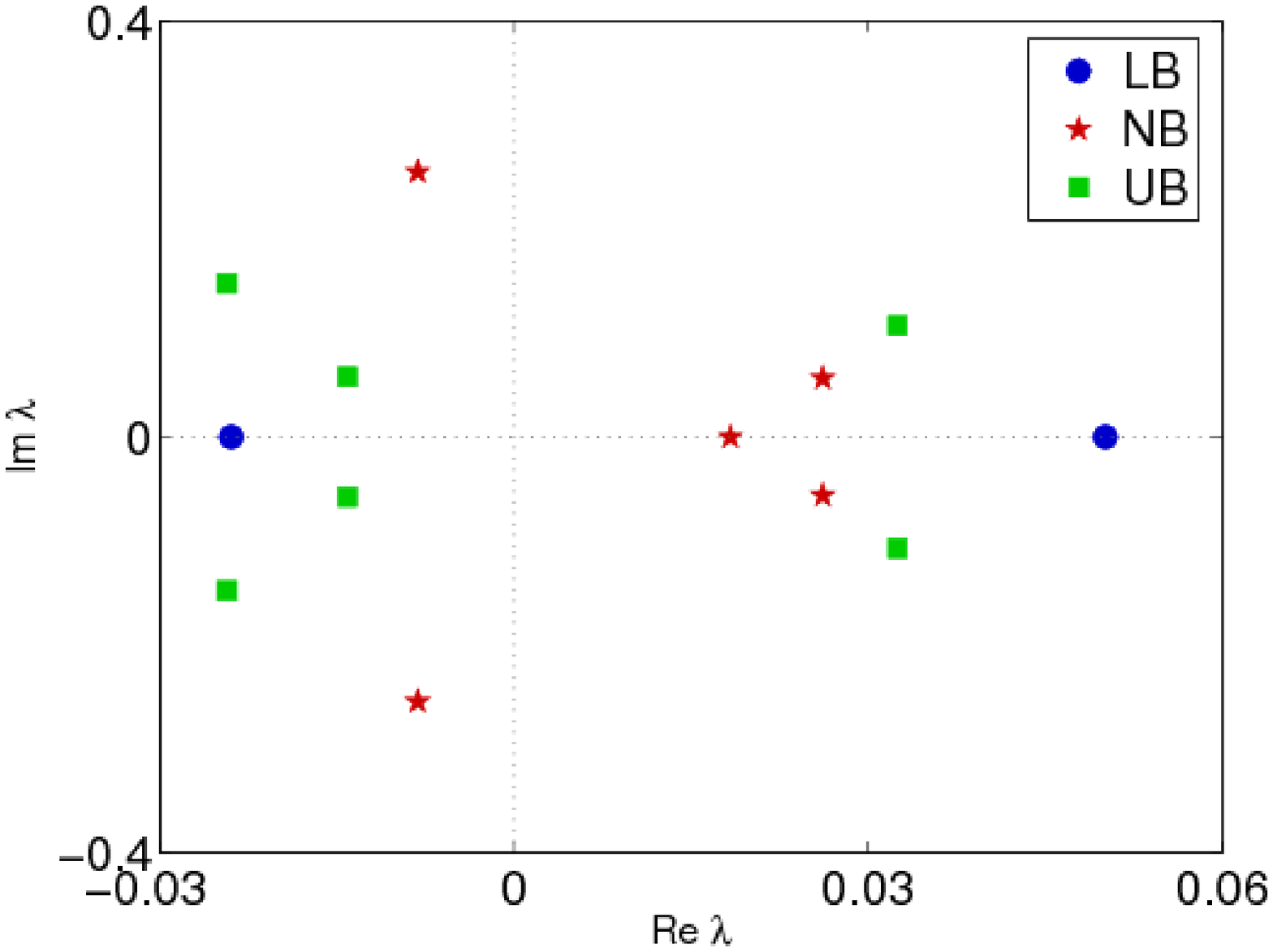}
     }
%\vspace*{-5pt}
\caption{
Leading \uLB, \uNB, \uUB\ eigenvalues in (a) the full space $\bbU$
(b) the $S$-invariant subspace $\bbU_S$. Numerical values are listed in
tables \ref{t:LBlambda} and \ref{t:NBUBlambda}.
}
\label{f:UBLBNBlambda}
\end{figure}

Dynamics in the neighborhood of an \eqb\ solution are governed
by the linear stability matrix
\beq
[D\!F]_{mn} = \frac{\partial F_m}{\partial \ucoeff_n}
\eeq
\refFig{f:UBLBNBlambda} shows the leading (most unstable) eigenvalues
of \uLB, \uNB, and \uUB, computed with Arnoldi iteration
(\cite{Visw07b}). \refFig{f:UBLBNBlambda}\textit{(a)} shows all computed
eigenvalues; \reffig{f:UBLBNBlambda}\textit{(b)} shows those within the
$S$-invariant subspace $\bbU_S$. The eigenfunctions $\bv$ of $D\!F$
at the {\eqba} are either symmetric, $s \bv = \bv$, or antisymmetric,
$s \bv = -\bv$, as all
$s \in S$ are idempotent, $s^2 = 1$.  Thus, in general, the
dynamics carries small perturbations of these
\eqba\ into the full space $\bbU$. The $\uLB$ \eqb\ has a single
unstable eigenvalue (\cite{WGW07}). Within $\bbU_S$, the $\uUB$ {\eqb}
has a single unstable complex pair, and $\uNB$ has one unstable real
eigenvalue and one unstable complex pair.

The Arnoldi eigenvalues are accurate to $10^{-6}$, as
determined by repeated calculations with different random initial vectors,
and comparison of Arnoldi computations to analytically known eigenvalues of
the laminar {\eqb}. This level of accuracy results
from our use of off-center finite-differencing to estimate differentials of
the flow in the Arnoldi iteration: $Df^t |_u v = (f^t(u + \epsilon v)
- f^t(u))/\epsilon + O(\epsilon)$, with $\epsilon = 10^{-7}$.
Tables of numerical eigenvalues and their symmetries are given in 
\refsect{s:appeTables} (table \ref{t:LBlambda} and \ref{t:NBUBlambda})
and at {\tt channelflow.org}.

\subsection{Linearized evolution}

Let $\lambda, v_{\stagn}$ be an eigenvalue, eigenvector solution of
$D\!F|_{\ucoeff_{\stagn}}  v = \lambda \, v$
at the \eqb\ $\ucoeff_{\stagn}$.%
\footnote{ We indicate particular invariant solutions with subscripts,
such as $\ucoeff_{\text{\tiny LB}}$ or \uLB\ for the lower-branch \eqb\
solution. The $n$th eigenvalue is $\eigExp[n], \; n=1,2,\ldots$, in
order of decreasing real part. Whenever the context allows it, we shall
omit the eigenvalue and/or solution labels.}
Then the linearized \stateDsp\ dynamics
$\dot{v} = D\!F|_{\ucoeff_{\stagn}} v$ about $\ucoeff_{\stagn}$ has solution
$v(t) = e^{\Lyap t} v_{\stagn}$, and the initial condition
$\ucoeff(0) = \ucoeff_\stagn + \epsilon \, v_{\stagn}$ with
$\epsilon \, |v_{\stagn}| \ll 1$ evolves as
\beq
\ucoeff(t) = \ucoeff_\stagn + \epsilon \, v_{\stagn} \, e^{\Lyap t}
 + O(\epsilon^2)
 \,.
\ee{a_unstbReal}
The linearized evolution of the velocity field $\bu(\bx,t)$ can then be
derived by reconstructing the velocity fields from the corresponding
\stateDsp\ vectors, as discussed in \refsect{s:finite_rep}. Small perturbations
about $\bu_{\stagn}$ along the eigenfunction $\bv_{\stagn}$ evolve as
\beq
\bu(\bx, t) = \bu_\stagn(\bx) + \epsilon \, \bv_{\stagn}(\bx) \, e^{\Lyap t}
        + O(\epsilon^2)  \,.
\ee{u_unstbReal}

%%%%%%%%%%%%%%%%%%%%%%%%%%%%%%%%%%%%%%%%%%%%%%%%%%%%%%%%%%%%%%%%%%
%%%%%%%%%%%%%%%%%%%%%%%%%%%%%%%%%%%%%%%%%%%%%%%%%%%%%%%%%%%%%%%%%%

Complex eigenvalues and eigenvectors must be recast in real-valued form
prior to conversion to velocity fields, since each element of the {\stateDsp}
vector $v$ is the real or imaginary part of a complex-valued spectral
coefficient in a CFD expansion such as \refeq{CFDexpansion}.
Let $\lambda_{\tny{EQ}}^{(n,n+1)} = \eigRe^{(n)} \pm i \eigIm^{(n)}$ be a complex
eigenvalue pair and $v_{\tny{EQ}}^{(n,n+1)} = v_r^{(n)} \pm i v_i^{(n)}$ the
corresponding complex eigenvectors. Then (dropping superscripts) the initial
condition $\ucoeff(0) = \ucoeff_{\tny{EQ}} + \epsilon \, v_r$ evolves as a
real-valued spiral
\beq
\ucoeff(t) = \ucoeff_{\tny{EQ}} + \epsilon \,
          (v_r \cos \eigIm t - v_i \sin \eigIm t) \,
          e^{\eigRe t} + O(\epsilon^2) \,.
\label{a_unstbCmplx}
\eeq
Real-valued fields $\bv_r$ and $\bv_i$ can be reconstructed from the
real-valued vectors $v_r$ and $v_i$, and the real-valued initial
velocity field $\bu(0) = \bu_{\tny{EQ}} + \epsilon \, \bv_r$ evolves as
\beq
\bu(t) = \bu_{\tny{EQ}} + \epsilon \,
(\bv_r \cos \eigIm t - \bv_i \sin \eigIm t) \, e^{\eigRe t}
 + O(\epsilon^2) \,.
%\label{cmplx_linear_u_soln}
\label{u_unstbCmplx}
\eeq
%\refFig{f:NBef01} shows $\vNB^{\pm}$ for the complex
%pairs $\lambda_{\tny{NB}}^{(2,3)}$ and $\vNB^{(2,3)}$ of \uNB.

\subsection{Unstable manifolds}
\label{s:unstable_manifolds}

Let $\Wmnfld{s}{EQ}$ ($\Wmnfld{u}{EQ}$) denote the stable
(unstable) manifold of \eqb\ $\uEQ$. For each real-valued unstable
eigenvalue $\Lyap^{(n)}$, we shall refer to the orbit of an
infinitesimal perturbation of \uEQ\ along the corresponding
eigenfunction $\vEQ^{(n)}$ as $\Wmnfld{u (n)}{EQ}$.
This part of the \uEQ\ unstable manifold is 1-dimensional and can be
computed by DNS integration of the initial conditions
$\uEQ \pm \epsilon \, \vEQ^{(n)}$, where $\epsilon \ll 1$.

For an unstable complex pair $\lambda^{(n,n+1)}$ of \eqb\ $\uEQ$,
let $\Wmnfld{u (n,n+1)}{EQ}$ denote the orbit of a {circle}
 of infinitesimal radius in the plane about \uEQ\ spanned
by $\bv^{(n)}_{r}, \bv^{(n)}_{i}$. This part of the \uEQ\ unstable manifold is 2-dimensional;
its shape can be traced out by computing a set of trajectories with initial
conditions $\uEQ + \epsilon (\bv_{r}^{(n)} \cos \theta + \bv_{i}^{(n)} \sin \theta)$
for a set of values of $\theta$. In practice, one obtains a
more uniform
distribution of trajectories by setting initial conditions along the line
$\uEQ + \epsilon \, \bv_{r}^{(n)}$, for a set of values of $\epsilon$.

The global unstable manifolds $\Wmnfld{u (n)}{EQ}$ and
$\Wmnfld{u (n,n+1)}{EQ}$ are invariant sets that
preserve the symmetries shared by the \eqb\ and the eigenvectors
from which they are generated.
The $S$-invariant subspace portions of the unstable manifolds of
\uLB, \uUB, and \uNB\ have dimensionality of 1, 2, and 3, respectively,
see \reffig{f:UBLBNBlambda}\textit{(b)}.
In what follows, we will focus on these low-dimension unstable
manifolds confined to the $\bbU_S$ subspace.

%% file: geometry.tex
% geometry.tex
% $Author: gibson $ $Date: 2008-02-11 15:11:56 +0530 (Mon, 11 Feb 2008) $

% \section{A tour of the plane Couette \statesp}
% \label{s:geometry}

We now turn to the main theme of this paper:
{\em exact {\stateDsp} portraiture} of
\pCf\ dynamics. The \stateDsp\ portraits are
{\em dynamically intrinsic}, since the projections
are defined in terms of solutions of the equations of
motion, and {\em representation independent}, since the projection
operation (the inner product \refeq{innerproduct}) is independent
of the numerical representation.
The method is by no means restricted to \pCf\ or
our choice of \stateDsp\ representation or CFD algorithm.
It can be applied to any high-dimensional dissipative flow, for
example the Kuramoto-Sivashinsky flow (\cite{SCD07}).
Production of \stateDsp\ portraits requires numerically
computed physical states (such as \eqb\ solutions and their linear
stability eigenfunctions), an algorithm for integrating dynamics,
and a method of computing the inner product between states
over the physical domain.

\subsection{Peering into $\infty$-dimensional \statesp s}
\label{s:vision}

Numerical methods have advanced to the point where it is possible
to compute highly accurate unstable exact \cohStr s in low-Reynolds
shear flows. How is one to visualize them?
Even though fully-resolved solutions of \NSe\
are embedded in $10^5$ or higher dimensional \statesp s,
there are few unstable eigendirections
for \Reynolds\ close to the onset of turbulence.
The associated asymptotic
strange attractors / repellers might thus be amenable to
dynamical systems visualizations, such as trajectory projections,
Poincar\'e sections, \stateDsp\ partitions
and symbolic dynamics description.

In this section, we show that revealing, representation-independent
\stateDsp\ portraits can be defined in terms of invariants of the
dynamical system. The idea is to choose as a basis set states of the 
fluid with characteristics of recurrent coherent structures, and to 
project the evolving fluid state $\bu(t)$ onto this basis with the
energy norm \refeq{InDrag} inner product
\begin{align}
 (\bu, \bv)  &= \frac{1}{V}
                \int_\Omega \! d \bx \;
                       \bu \cdot \bv \,,
       \quad \Norm{\bu}^2  = (\bu, \bu) \,.
\label{innerproduct}
\end{align}
That is, we form orthonormal basis functions
$\{\ben{1}$, $\ben{2}$, $\ldots, \ben{n}\}$ from
a set of linearly independent fluid states and
produce a \stateDsp\ trajectory
\beq
\ssp(t) =(\sspn{1}, \sspn{2}, \cdots, \sspn{n}, \cdots)(t)
    \,,\qquad
\sspn{n}(t) = (\bu(t), \ben{n})
\ee{intrSspTraj}
in the $\{\ben{n}\}$ coordinate frame by \refeq{innerproduct}.
The projection can be viewed in any of the $2d$ planes
$\{\ben{m}, \ben{n}\}$ or in $3d$ perspective views
$\{\ben{\ell},\ben{m}, \ben{n}\}$. The resulting
portraits depend on the fluid states involved
and not on the choice of numerical representation. Orthonormality
of the basis set is not strictly necessary, but with it, distances
are directly related to  \refeq{InDrag}, the energy norm of $\bu$.

The low-dimensional projections presented in this section are closely related to
other finite approaches to Navier-Stokes in both technical methods and
purpose. For example, on a technical level, the projections in this section
differ from the finite discretizations discussed in \refsect{s:finite_rep}
only by degree of dimensionality. If the dimension $n$ of the {\stateDsp}
representation \refeq{intrSspTraj} were taken to the dimension $d$ of
the fully-resolved numerical discretization, the two discretizations would
be related by a simple linear transformation. We emphasize the differences
between the two through notation: $\ucoeff$ for the high-dimensional vector
of coefficients of a fully-resolved numerical discretization, and $\ssp$ for
the low-dimensional coordinates of a \stateDsp\ portrait.

The projection methods here are quite similar in spirit to the
low-dimensional projections of the \cite{Aubry88} POD modeling approach,
in that they
aim to capture key features and dynamics of the system in just a few
dimensions. Indeed, our use of the $L^2$ inner product, orthonormal
basis functions, and the very idea of constructing a basis from
characteristic states derive directly from POD modeling. But the
methods presented here depart from the POD in two key points: (1) We
construct basis sets from {\em exact solutions of the full-resolved
dynamics} rather than from the empirical eigenfunctions of the POD.
Exact solutions and their linear stability modes
(a) characterize coherent fluid states precisely,
compared to the truncated expansions of the POD, (b) allow for different
basis sets and projections for different purposes and different regions
of \statesp, and (c) are not limited to Fourier modes and $O(2)$ symmetry
in homogeneous directions.
(2) We deploy low-dimensional {\em visualization} without any low-dimensional
{\em modeling}. The dynamics are computed with fully-resolved direct
numerical simulations and projected onto basis sets to produce
low-dimensional \stateDsp\ portraits,
tailored to specific purposes and specific regions of \statesp.
The portraits reveal dynamical information visually, providing
insight to dynamics that can guide further analysis. Specifically, we
do not suggest that any of our low-dimensional projections is
suited to a global projection of the \stateDsp\ dynamics into a
low-dimensional ODE model.

\subsection{A global basis spanned by discrete translations of \uUB}
\label{s:UBglobal}

There is an infinity of possible basis sets, but two choices appear particularly
natural:
(a) global basis sets, determined by a set of dynamically important and distinct
states, or
(b) local basis sets, defined in terms of a given \eqb\ \uEQ\ and its linear
stability eigenfunctions $\vEQ^{(n)}$.
An example of a local coordinate system based on eigenfunctions of the \uUB\
{\eqb} is presented in \refsect{s:UBunstManf}; an example of a global
basis is defined here and used to construct \stateDsp\ portraits in
\refsect{s:globalPCf}.

The projection for a global \stateDsp\ portrait should emphasize important
global features of the flow.
For example, for a system with three distinct {\eqba}, a good first guess
for a plane of projection would be the plane containing the three {\eqba}.
The system under study has three distinct {\eqba} \uUB, \uLB, and \uNB,
each appearing in four spatial phases, plus the laminar {\eqb} at the
origin. We have found that for the $S$-invariant subspace $\bbU_S$ 
the irreducible representations of the half-cell
translations group $T$ (\refsect{s:PCFsymm}) provide natural linear
combinations of a given \eqb\ and its translations.

For example, a set of orthonormal basis functions based on \uUB\ and its half-cell
translated siblings can be generated by the four irreducible representations
of the $D_2$ dihedral group $T = \{1,\tau_x,\tau_z,\tau_{xz}\}$
(see \refsect{s:PCFsymm}):
\begin{align}
 & \qquad\qquad\qquad\qquad\qquad\qquad
              ~~~~ \tau_x ~~ \tau_z  ~~  \tau_{xz}
    \nnu\\
\beUBg{1} &= c_1 (1 + \tau_x + \tau_z + \tau_{xz})
      \, \uUB      ~~~~  S  ~~  S   ~~   S
    \nnu\\
\beUBg{2} &= c_2 (1 + \tau_x - \tau_z - \tau_{xz})
      \, \uUB     ~~~~  S  ~~  A   ~~   A
    \nnu\\
\beUBg{3} &= c_3 (1 - \tau_x + \tau_z - \tau_{xz})
      \, \uUB      ~~~~  A  ~~  S   ~~   A
     \label{ek_defn}\\
\beUBg{4} &= c_4 (1 - \tau_x - \tau_z + \tau_{xz})
      \, \uUB      ~~~~  A  ~~  A   ~~   S
\,.
    \nnu
\end{align}
where $c_n$ is a normalization constant determined by $\Norm{\beUBg{n}} = 1$.
% Defining $c_n$ in terms of its length along $\be_n$ is more geometric, but 
% I think requires more brain cycles to understand. 
The last 3 columns indicate the symmetry of each basis function under half-cell
translations; e.g.\ $S$ in the $\tau_x$ column implies that
$\tau_x \beUBg{n} = \beUBg{n}$ and an $A$ that $\tau_x \beUBg{n} = - \beUBg{n}$.
As the `velocity'  $\bu$ in the \NSe\ \refeq{NavStokesDev}
for \pCf\ is the difference from laminar flow, the origin in {\stateDsp}
portraits corresponds to the laminar \eqb\ \uLM. This origin is shared by all
symmetry-invariant subspaces, as $\uLM = 0$ is invariant under all symmetries
of the flow. Note, however, that the basis functions $\beUBg{n}$ are not themselves
invariant solutions of Navier-Stokes; rather, they form an orthogonal coordinate
system that spans the four translations of $\uUB$ within the $S$-invariant
subspace $\bbUsymm$.

The evolution of a state $\bu \in \bbUsymm$ is represented in this projection
by the trajectory $\sspg{}(t) = (\sspg{1}, \sspg{2}, \sspg{3}, \sspg{4})(t)$
with $\sspg{n}(t) = (\bu(t), \beUBg{n})$. As discussed in \refsect{s:vision},
this is a low-dimensional projection intended for visualization. The dimensionality
is lower than the full \statesp, so trajectories can appear to cross
in such projections.
We emphasize again that this is one of many possible projections
that can be constructed from linear combinations of exact solutions, their spatial
translations, and their eigenfunctions. An example of a more complex basis construction
is given in \refsect{s:UBunstManf}. 
%To keep things simple, for the remainder of
%this tour we focus on dynamics within the $S$-invariant  subspace $\bbUsymm$.
% Said this earlier

\subsection{A global stroll through plane Couette \statesp}
\label{s:globalPCf}

\begin{figure}
\centering
\includegraphics[height=3in]{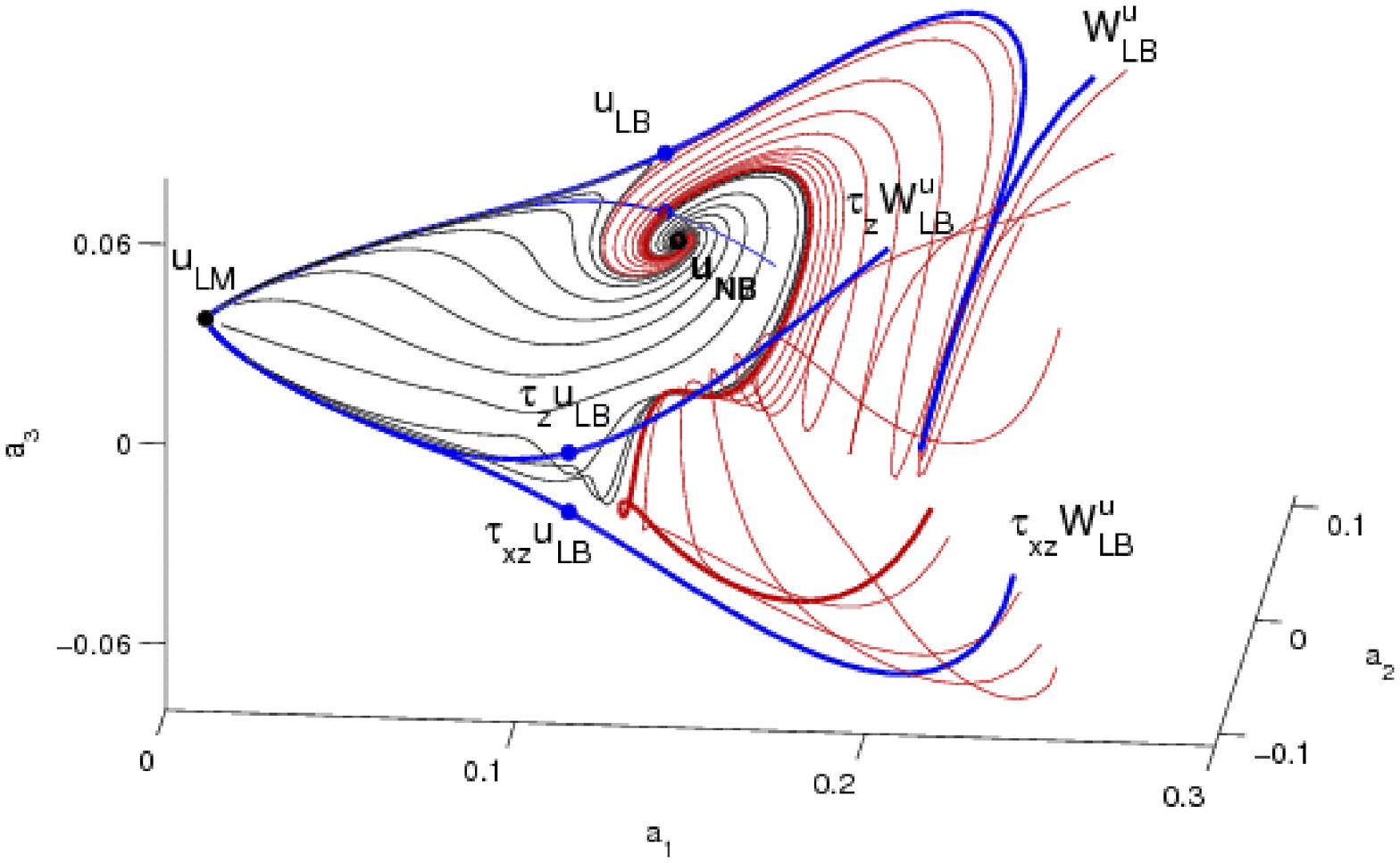}
\caption{
The unstable manifold of \uLB\ and its half-cell translations
$W^{u}_{\tny{LB}}$, $\tau_{xz} W^{u}_{\tny{LB}}$, etc. (thick \colorcomm{blue}{black}
lines) and a $2d$ portion of the \uNB\ unstable manifold
$W^{u (1,2)}_{\tny{NB}}$ (thin black and \colorcomm{red}{gray} lines).
$\uNB$ is shown in only one translation (center of spiral); all four
translations of $\uLB$ are shown (solid dots --the unlabeled dot
underneath $\uLB$ is $\tau_{x} \uLB$).
The thick \colorcomm{red}{gray} line makes the closest pass to 
$\tau_{xz} \uLB$ of the trajectories shown. The
projection is from 61,506 dimensions to 3 in the translation-symmetric
global basis $\{\beUBg{1}, \beUBg{2}, \beUBg{3}\}$ defined by \refeq{ek_defn}.
% \Wmnfld macro is broken within fig captions. jfg 2008-01-12
        }
\label{f:LBNBUB_unstable_b}
\end{figure}

With this road map in hand, let us take a stroll through the
\statesp\ of a transiently turbulent \pCf .
Like  many dynamical narratives,
this will be a long walk through unfamiliar landscape with many landmarks
of local interest. We undertake the tour for
several reasons. The main message is
that now such a promenade is possible even in $10^5$ dimensions.
But a detailed road map is a necessary prerequisite
for solving
at least three outstanding problems:
(a) uncovering the interrelations between (in principle infinite number of)
invariant solutions, such as those of \reffig{f:LBNBUB_unstable_b},
(b) a partition of \statesp\ is a needed for a systematic exploration
of dynamical invariant structures such as \rpo s, and
(c) explicit linear stability \ev s and their unstable-manifold continuations
will be needed to control and chaperone a given fluid state to
a desired target state.

\begin{figure}
\centering
\includegraphics[height=3in]{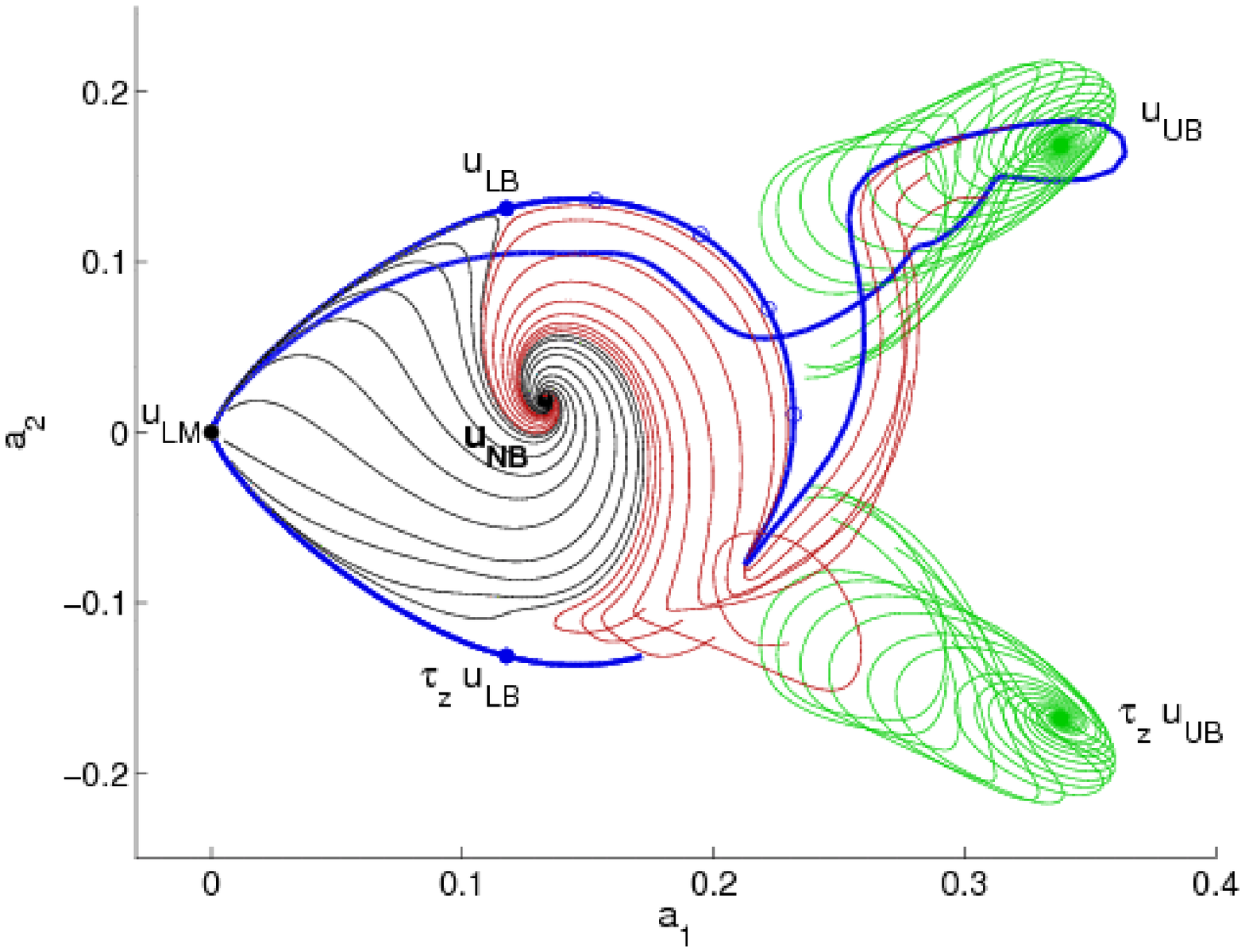}
\caption{
A {\statesp} portrait of \pCf\ for $\Reynolds = 400$ and
$[L_x,L_y,L_z] = $ $[2\upi/1.14,2,4\upi/5]$, projected from $61\,506$
dimensions to 2. The labeled points are exact {\eqb} (steady-state)
solutions of the \NSe\ (see \refsect{s:equilibria}); the curved trajectories
are fully-resolved time-dependent numerical integrations of \NS\ projected
onto the $(\beUBg{1},\beUBg{2})$  plane defined by \refeq{ek_defn}.
$W^{u}_{\tny{LB}}$, the $1d$ unstable manifold of the `\lbranch' \eqb\ \uLB, and
$\tau_z W^{u}_{\tny{LB}}$, its half-cell translation in $z$, are shown with thick
\colorcomm{blue}{black} lines. $W^{u (1,2)}_{\tny{NB}}$, a $2d$ portion of the unstable manifold of
the `newbie' \eqb\ \uNB, is shown with thin black and \colorcomm{red}{gray} spirals emanating
from \uNB. Similarly, the \colorcomm{thin green}{dark gray} lines spirally out of $\uUB$ and $\tau_z \uUB$
indicate $W^{u, S}_{\tny{UB}}$ and $\tau_z W^{u, S}_{\tny{UB}}$, the $2d$ unstable
manifolds of \uUB\ and its half-cell translation $\tau_z \uUB$ within the
$S$-invariant subspace $\bbUsymm$. Open dots along $W^{u}_{\tny{LB}}$ show initial 
conditions for Newton-GMRES searches used to find \uNB.
The plane of the projection is defined in terms of the {\eqb} solutions;
it is {\em dynamically invariant} and independent of the numerical
representation. See \refsect{s:UBglobal} and \refsect{s:globalPCf} for discussions
of the projection and the dynamics.}
\label{f:LBNBUB_portrait_a}
\end{figure}

Our first example of a global {\stateDsp} portrait of \pCf\ is
\reffig{f:LBNBUB_portrait_a}. Here trajectories in the unstable
manifolds of \uLB, \uNB, \uUB\ and several of their half-cell
translations are projected onto $\{\beUBg{1}, \beUBg{2}\}$ plane defined by
\refeq{ek_defn}. Both $\beUBg{1}$ and $\beUBg{2}$
are symmetric in $\tau_x$, so points related by half-cell
translations in $x$ (such as \uLB\ and $\tau_x \uLB$) map to the
same point in this projection.
The basis function $\beUBg{2}$ is antisymmetric in $\tau_z$,
so half-cell translations in $z$ appear symmetrically opposite along
$\sspg{2}$.
\uNB\ and its unstable manifold are shown only in a single $z$ translation,
since the reversed orientation of the unstable spiral of $\tau_z \uNB$
only clutters the picture.

{\bf The \uLB\ unstable manifold} \Wmnfld{u}{LB}\ forms the backbone of
the dynamics shown in \refFig{f:LBNBUB_portrait_a}.
The \uLB\ \eqb\ has a single real-valued unstable eigenvalue, as shown
in \reffig{f:UBLBNBlambda} for $\Reynolds = 400$ and for much higher
$\Reynolds$ in \cite{WGW07}. \Wmnfld{u}{LB}\ is therefore $1d$ and
can be computed
in its entirety as discussed in \refsect{s:unstable_manifolds}.
\refFig{f:LBNBUB_portrait_a} shows the projection of {\Wmnfld{u}{LB}}
onto $\{\beUBg{1}, \beUBg{2}\}$ plotted with thick \colorcomm{blue}{black} lines.
Both branches of \Wmnfld{u}{LB} decay to laminar flow, one immediately,
and the other after a turbulent excursion towards \uUB.
The portion of the unstable manifold of $\tau_z \uLB$ shown here
was obtained by applying
the $\tau_z$  shift, $\ssp_2 \to - \ssp_2$, to \Wmnfld{u}{LB}.

We were lead to the discovery of the `newbie' {\eqb} \uNB\ by the circular
curvature of \Wmnfld{u}{LB}\ and $\tau_z \Wmnfld{u}{LB}$ in the projection
of \reffig{f:LBNBUB_portrait_a}, which suggested the possibility of an {\eqb}
with a complex eigenvalue near the center of curvature. We initiated
Newton-GMRES searches for an {\eqb} at several positions between noon and
three o'clock along \Wmnfld{u}{LB}, as pictured in \reffig{f:LBNBUB_portrait_a};
each search converged either on \uLB\ or on the new {\eqb} \uNB.

{\bf The \uNB\ unstable manifold} \Wmnfld{u}{NB}:
Within $\bbUsymm$, the \uNB\ \eqb\ has a complex pair of unstable
eigenvalues and one real unstable eigenvalue  (\reffig{f:UBLBNBlambda}\textit{(b)}).
The instability of the real eigenvalue is weaker than the complex pair;
we omit it from consideration here and focus on the $2d$
subset \Wmnfld{u (1,2)}{NB}\ corresponding to the complex pair
$\lambda_{\tny{NB}}^{(1,2)}$ with eigenvectors $v_{\tny{NB}}^{(1,2)}$.
\Wmnfld{u (1,2)}{NB}\ is shown in \reffig{f:LBNBUB_portrait_a} as a spiral
of trajectories emanating from \uNB, calculated as discussed in
\refsect{s:unstable_manifolds}. This simple geometric picture produces our
first striking result: the $2d$ surface \Wmnfld{u (1,2)}{NB}\ is apparently
bounded by the $1d$ curve \Wmnfld{u}{LB}.

{\bf A heteroclinic connection from \uNB\ to \uLB}:
As it approaches \uLB, \Wmnfld{u (1,2)}{NB}\ separates along the two
branches of \Wmnfld{u}{LB}. Since \uLB\ has a single unstable
eigenvalue, we expect that a single trajectory
in \Wmnfld{u (1,2)}{NB}\ straddles the split along
\Wmnfld{u}{LB}\ and
is drawn in towards \uLB\ along its stable \ev s as
$t \rightarrow \infty$, forming a heteroclinic
connection from \uNB\ to \uLB.

This is a strikingly unexpected result. In dimensions higher than two,
heteroclinic connections are nongeneric, since it is unusual that a
$1d$ trajectory can be arranged to strike a particular
zero-dimensional point. However, discrete symmetries and the
dimensionality of the $\uLB$ unstable manifold make
heteroclinic connections possible in this case (\cite{KNSks90,Holmes96,SCD07}).
The set of candidate trajectories emerging from the neighborhood of
\uNB\ is increased from one dimension to two by the complex instability
(or three if $\lambda_{\tny{NB}}^{(3)}$ is considered as well). The
dimensionality of \statesp\ near the target \uLB\ is effectively
reduced to one by its codimension-1 set of stable eigenvalues.

Considered in the full space $\bbU$, the continuous
translation symmetry increases the dimensionality of both the
candidate trajectories and the target by two. However, the invariance
of $\bbU_S$ under Navier-Stokes immediately restricts possible
heteroclinic connections between the torus of \uNB\ and {\uLB}
translations to the four translations of \uLB\ within $\bbU_S$:
if a trajectory in the unstable manifold of {\uNB} terminates at a
$\uLB$ state, it may do so only at $\uLB, \tau_x \uLB,
\tau_z \uLB,$ or $\tau_{xz} \uLB$.
Note also that most weakly stable eigenvalues of \uLB ,
$\lambda_{\tny{LB}}^{(4)}$ through $\lambda_{\tny{LB}}^{(8)}$, are
outside the $\bbU_S$ subspace, so trajectories in
\Wmnfld{u (1,2)}{NB}\ are
forced to approach \uLB\ along the more strongly contracting
eigendirections of $\lambda_{\tny{LB}}^{(9)}$ and
$\lambda_{\tny{LB}}^{(10)}$ (\reftab{t:LBlambda}).

The heteroclinic connection from \uNB\ to \uLB\ forms a boundary between trajectories that
decay immediately to laminar flow and those that grow towards transient
turbulence. Those that pass near \uLB\ and grow to turbulence follow
the unstable manifold of \uLB\ into a region near the \uUB\ \eqb.
For $\Reynolds = 400$ and $[L_x,L_y,L_z] = $ $[2\upi/1.14,2,4\upi/5]$,
all generic initial conditions investigated so far
ultimately decay to laminar.  But, at higher Reynolds
numbers and larger aspect ratios for which turbulence is sustained,
we expect that the $\uNB \rightarrow \uLB$ heteroclinic connection will form a
$1d$ portion of the boundary of the
laminar state's basin of attraction. This $1d$ boundary should be
extendable to $2d$ by adding the third unstable eigenvalue of
\uNB\ into consideration.

Lastly, we note that it is not possible to determine from
\reffig{f:LBNBUB_portrait_a} alone whether the heteroclinic connection
from $\uNB$ goes to $\uLB$ or $\tau_x \uLB$, since both of these map
to the same point in the $\{\beUBg{1}, \beUBg{2}\}$ plane of projection.
\refFig{f:LBNBUB_unstable_b}
(discussed below) resolves this question and shows that the connection is
indeed from \uNB\ to \uLB.

%%%%%%%%%%%%%%%%%%%%%%%%%%%%%%%%%%%%%%%%%%%%%%%%%%%%%%%%%%%%%%
%%%%%%%%%%%%%%%%%%%%%%%%%%%%%%%%%%%%%%%%%%%%%%%%%%%%%%%%%%%%%%

{\bf Dynamics near $\tau_{z} \uLB$ and $\tau_{xz} \uLB$}:
A second separation of \uNB's unstable manifold \Wmnfld{u (1,2)}{NB}\ occurs
in the bottom half of \reffig{f:LBNBUB_portrait_a}, near $\tau_z \uLB$.
Trajectories on the laminar side of $\tau_z \uLB$ 
follow its unstable manifold towards the laminar state; those on the
other side head towards turbulence in the direction of the
$\tau_z \uLB$ unstable manifold.

The dynamics in the region near $\tau_z \uLB$ can be clarified by adding a third coordinate
$\sspn{3} = (\bu, \beUBg{3})$ to the $2d$ projection of
\reffig{f:LBNBUB_portrait_a}.
Since $\beUBg{3}$ is antisymmetric in $\tau_x$, the $a_3$ coordinate
distinguishes states such as $\tau_{z} \uLB$ and $\tau_{xz} \uLB$ that
are related by a $\tau_x$ translation and so lie on top of each other
in the projection of \reffig{f:LBNBUB_portrait_a}:
$(\tau_{z} \uLB, \beUBg{3}) = - \, (\tau_{xz} \uLB, \beUBg{3})$.
\refFig{f:LBNBUB_unstable_b}
shows a $3d$ perspective of $(\sspg{1},\sspg{2},\sspg{3})$ which
reveals that the second separation of $W^{u (1,2)}_{\text{\tiny NB}}$,
%\Wmnfld{u (1,2)}{NB},
unlike the first, does not
result from a heteroclinic connection between \uNB\ and a translation
of $\uLB$. As trajectories straddling the split near $\tau_{z} \uLB$
and $\tau_{xz} \uLB$ are refined, they approach neither of these
points. Likewise, the extensions of the refined trajectories approach
neither the $W^{u}_{\tau_{z} \tny{LB}} = \tau_z W^{u}_{\tny{LB}}$
nor $W^{u}_{\tau_{xz} \tny{LB}} = \tau_{xz} W^{u}_{\tny{LB}}$
unstable manifolds. The thick \colorcomm{red}{gray} trajectory in 
\reffig{f:LBNBUB_unstable_b} passes closer to $\tau_{xz} \uLB$ than 
the other trajectories shown, but its recedes from 
$\tau_{xz} W^{u}_{\tny{LB}}$ instead of approaching it.

The geometry of \Wmnfld{u (1,2)}{NB}\ in this region is fairly complex.
The separation of trajectories between $\tau_z \uLB$ and $\tau_{xz} \uLB$
suggests that another {\eqb} might exist in this region; however,
our Newton-GMRES searches initiated in this region converged on $\tau_z \uLB$ or $\tau_{xz} \uLB$.
It is clear, however, that the geometry of \Wmnfld{u (1,2)}{NB} is shaped by
the unstable manifolds of \uLB\ and two of its translations, namely, $\Wmnfld{u}{LB}$,
$W^{u}_{\tau_{z} \text{\tiny LB}}$, and $W^{u}_{\tau_{xz} \text{\tiny LB}}$.
The upper-branch solution also plays a role: in \reffig{f:LBNBUB_portrait_a}
one trajectory within \Wmnfld{u (1,2)}{NB}\ is drawn towards
$\tau_z \uUB$ and follows trajectories in its unstable manifold.
The perspective of \reffig{f:LBNBUB_unstable_b} also identifies \uLB\
and not $\tau_x \uLB$ as the endpoint of the heteroclinic connection
discussed above.

Thus, with two simple \stateDsp\ portraits, we have identified several
regions in \statesp\ that trigger transitions toward qualitatively
different types of flow.
%(and within $\bbU_S$, each of these regions is
%multiplied by four by the discrete symmetries).
We expect that identification of such \stateDsp\ regions
will be extremely valuable in the development of nonlinear control
strategies for wall-bounded turbulence.

\subsection{A local \stateDsp\ portrait:
            the unstable manifold of \uUB\ in $\bbUsymm$}
\label{s:UBunstManf}

The eigenfunctions of an \eqb\ provide a natural coordinate system
for viewing its local dynamics. Within the $S$-invariant subspace $\bbUsymm$,
\uUB\ has a single complex pair of unstable eigenvalues, which define
a plane of local linear oscillation and two natural directions for a local
coordinate system. The $2d$ portion of \Wmnfld{u}{UB} within $\bbU_S$,
which we denote by \Wmnfld{u,S}{UB}, departs from this plane
as the distance from \uUB\ increases and the magnitudes of nonlinear terms
in the local Taylor expansion become nonnegligible. But since the nature of
this nonlinearity was unknown, it was not immediately clear in our
investigations how to choose a third basis function for a $3d$ projection
of local $\uUB$ dynamics. We tried a variety of candidates, including principal
components analysis (i.e. local POD) on numerically integrated trajectories as
they deviate from the plane of oscillation. This initial exploration
suggested that the dominant nonlinear effects about \uUB\ are in fact the
linearized dynamics around its half-cell translation $\tau_x \uUB$.

We then constructed a basis set by Gram-Schmidt orthogonalization of the
plane of oscillation $\bv^{(1)}_{r, \text{\tiny UB}}$, $\bv^{(1)}_{i, \text{\tiny UB}}$
of the unstable complex eigenvalue pair $\lambda_{\tny{UB}}^{(1,2)}$
(see \refsect{s:unstable_manifolds}) and $(\tau_x \uUB - \uUB)$,
that is, the direction between \uUB\ and its half-cell translation in $x$.
We indicate the Gram-Schmidt orthogonalized basis and coordinates with a
$\lambda$ superscript, $\{\beUBl{1}, \beUBl{2}, \beUBl{3}\}$ and
$\sspl{n}(t) = (\bu(t), \beUBl{n})$, to indicate its construction from
the unstable \uUB\ eigenfunctions and the $\uUB$ to $\tau_x \uUB$ line.

%%%%%%%%%%%%%%%%%%%%%%%%%%%%%%%%%%%%%%%%%%%%%%%%%%%%%%%%%%%%%%
\begin{figure}%[ht]
%\vspace*{-5pt}
\centering
{\small (a)} \includegraphics[height=0.23\textheight]{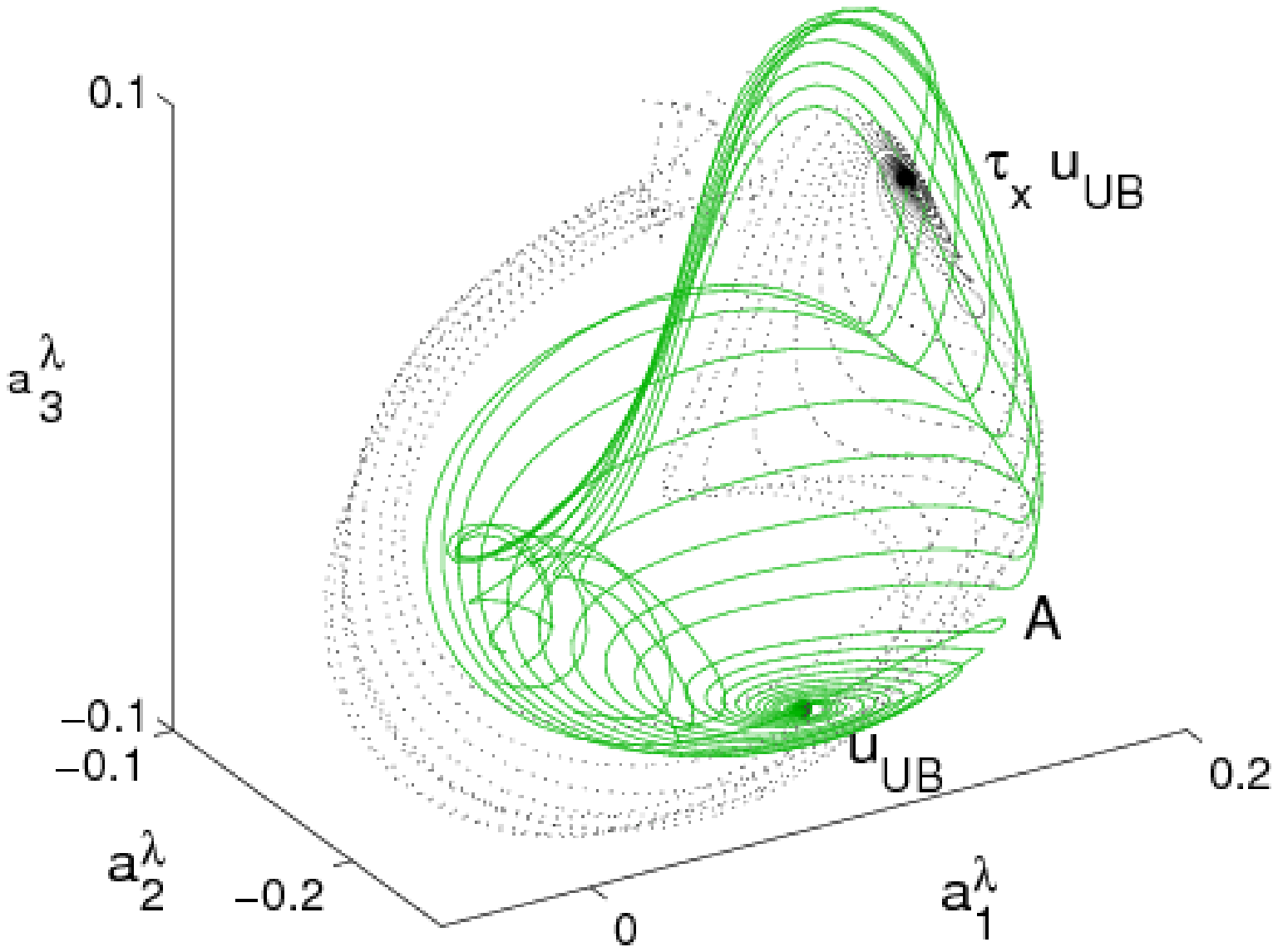}
~~~~
{\small (b)} \includegraphics[height=0.23\textheight]{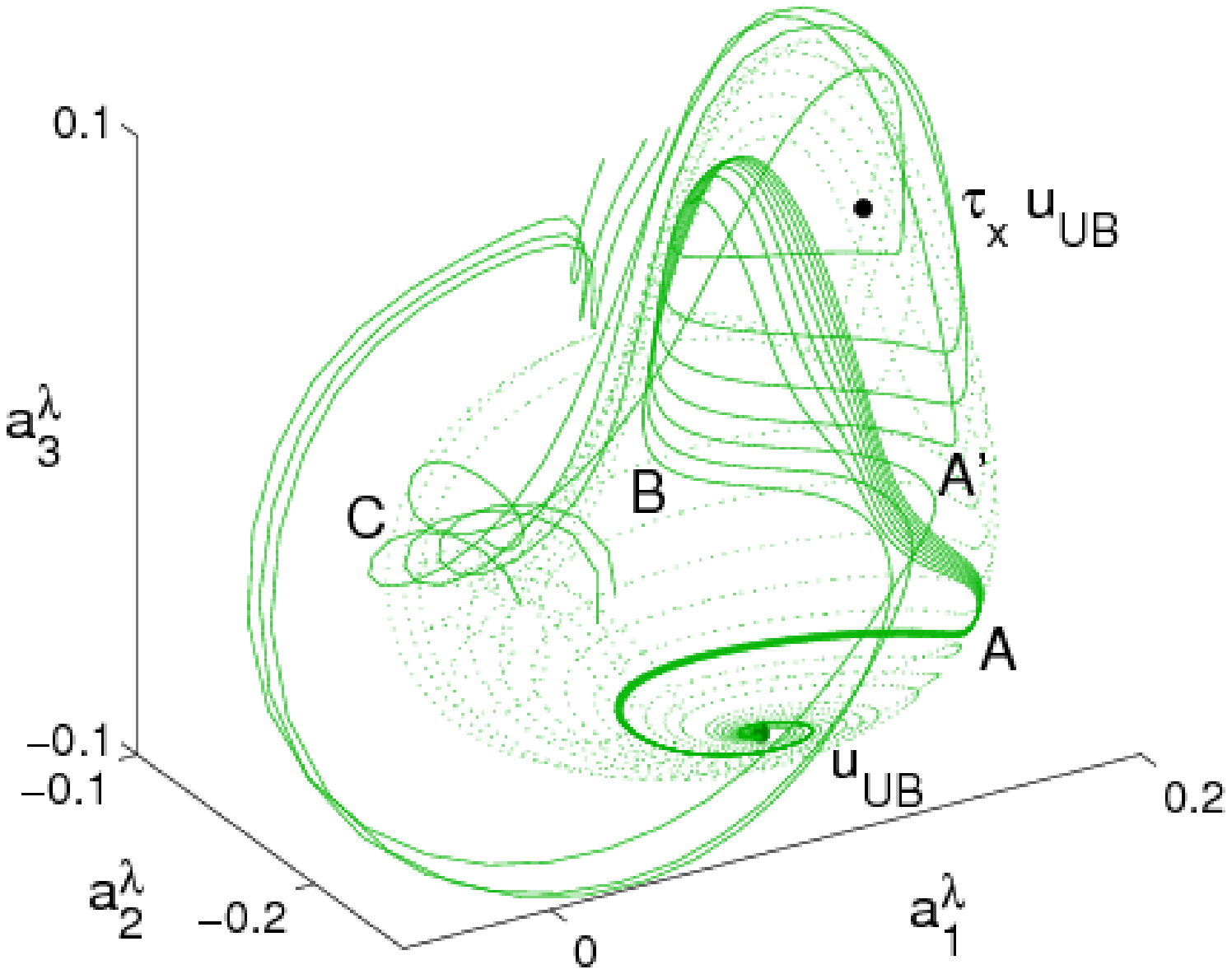}
\caption{
(a) The unstable manifolds $W^{u,S}_{\tny{UB}}$ (solid \colorcomm{green}{gray} lines) and
$\tau_x W^{u,S}_{\tny{UB}}$ (dotted black lines) of the upper-branch
{\eqb} \uUB\  and its half-cell translation $\tau_x \uUB$.
(b) A refined view of dynamics within $W^{u,S}_{\tny{UB}}$.
The coordinates $(\sspl{1},\sspl{2},\sspl{3})$ are projections onto the
basis set $\{\beUBl{1}, \beUBl{2}, \beUBl{3}\}$ that spans the plane of
unstable oscillation around \uUB\ and the direction between
$\uUB$ and $\tau_x \uUB$. See \refsect{s:UBunstManf}.
}
\label{f:UBdynamics1}
\end{figure}
%%%%%%%%%%%%%%%%%%%%%%%%%%%%%%%%%%%%%%%%%%%%%%%%%%%%%%%%%%%%%
%

\refFig{f:UBdynamics1} shows \Wmnfld{u,S}{UB} in the $\beUBl{n}$ local coordinate
system. In \reffig{f:UBdynamics1}\,(\textit{a}), \Wmnfld{u,S}{UB}\ spirals out
from the center \uUB\ in the plane $\{\beUBl{1}, \beUBl{2}\}$ spanning
$\bv^{(1)}_{r,\tny{UB}}$, $\bv^{(1)}_{i,\tny{UB}}$. Strong nonlinearity and
strong trajectory separation first occur near point $A$: below $A$,
trajectories continue the unstable linear oscillation for another
cycle; above, they begin oscillation around $\tau_x \uUB$, following paths
similar to trajectories in $\tau_x \Wmnfld{u}\uUB$.
\refFig{f:UBdynamics1}\,(\textit{b}) shows a refinement of trajectories
in \Wmnfld{u,S}{UB} on the upper side of the split at $A$.
Near point $B$, the refined trajectories undergo a second split from their
neighbors shown in \reffig{f:UBdynamics1}\,(\textit{a}), and a third split
among themselves at $A'$. This behavior and marked similarity to the trajectories
of $\tau_x \Wmnfld{u, S}{UB}$ in \reffig{f:UBdynamics1}\,(\textit{b}) suggests
that dynamics in this region consists of alternating oscillations
around a symmetric pair of unstable {\eqb}, in a manner reminiscent of
the Lorenz system, until escape.

%{\bf A global, translation-symmetric view of the \uUB\ and
%     $\tau_x \uUB$ unstable manifolds}:

%%%%%%%%%%%%%%%%%%%%%%%%%%%%%%%%%%%%%%%%%%%%%%%%%%%%%%%%%%%%%%
\begin{figure}%[ht]
%\vspace*{-5pt}
\centering
\includegraphics[width=0.8\textwidth]{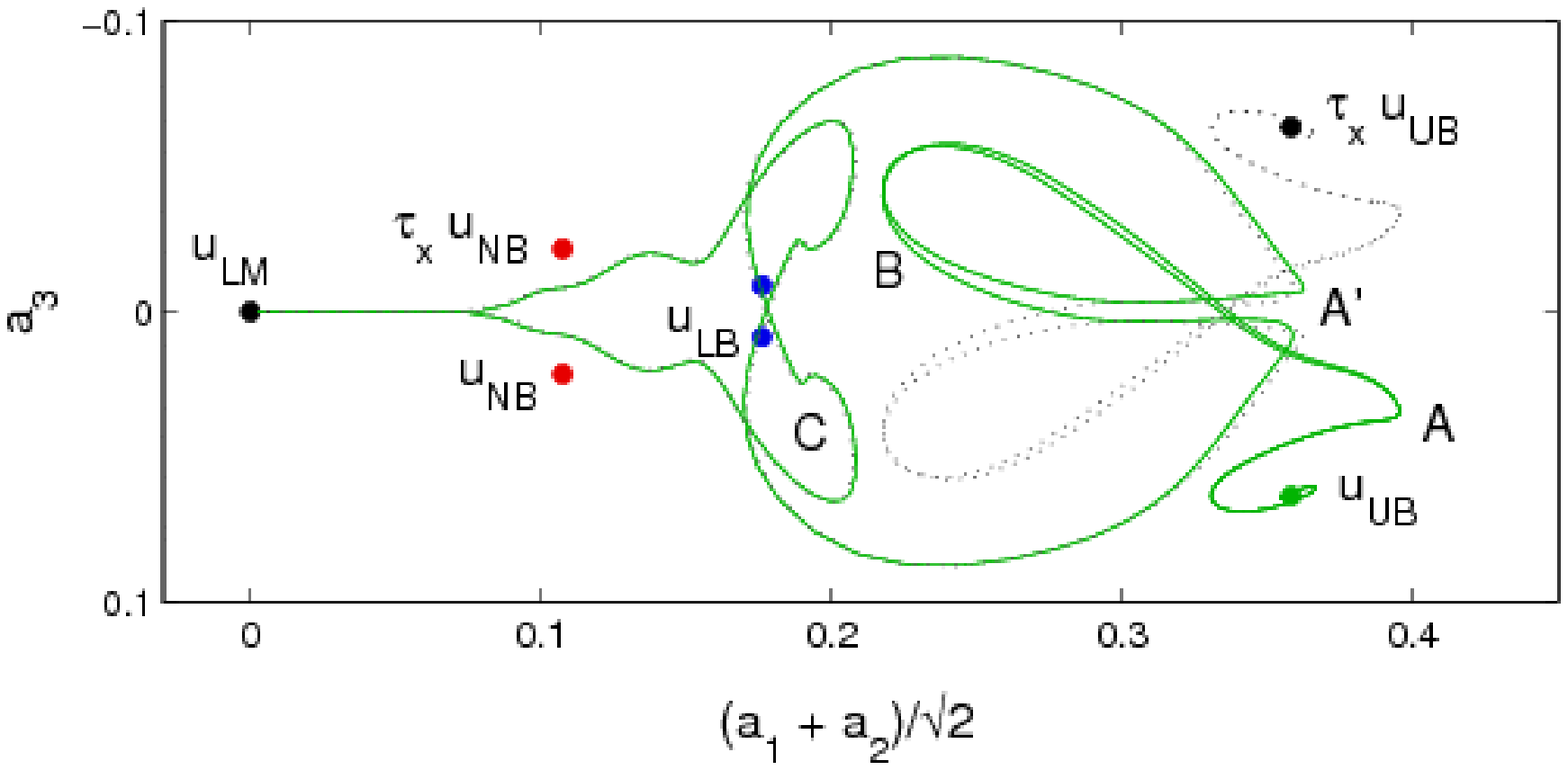}
\caption{
A pair of nearby trajectories in the unstable manifold of \uUB, from
\reffig{f:UBdynamics1}\textit{(a)} (solid, \colorcomm{green}{black}), plotted against their symmetric
counterparts in the unstable manifold of $\tau_x \uUB$ (dotted, black),
together with \uLB\ and $\tau_x \uLB$ (\colorcomm{blue}{black} dots, latter unlabeled),
\uNB\ and $\tau_x \uUB$ (\colorcomm{red}{gray} dots), and the laminar \eqb\ \uLM\ (black dot).
\uNB\ appears much closer to the laminar state than \uLB\ by an artifact of
the projection; see \reffig{f:LBNBUB_portrait_a} for another view.
The coordinates $\sspg{1}, \sspg{2}$ are in the global translation-symmetric
basis $\beUBg{1}, \beUBg{2}$ defined by \refeq{ek_defn} and used in
\reffig{f:LBNBUB_portrait_a} and \reffig{f:LBNBUB_unstable_b}.
}
\label{f:UBdynamics2}
\end{figure}
%%%%%%%%%%%%%%%%%%%%%%%%%%%%%%%%%%%%%%%%%%%%%%%%%%%%%%%%%%%%%%

This interpretation is reinforced by \reffig{f:UBdynamics2}, which shows the
pair of $\Wmnfld{u,S}{UB}$ trajectories from \reffig{f:UBdynamics1}\textit{(b)} that split at $A'$
together with their counterparts in $\tau_x \Wmnfld{u,S}{UB}$, replotted using
the global translational-symmetric basis \refeq{ek_defn}. The projection onto
$(\beUBg{1} + \beUBg{2})/\sqrt{2}$ and $\beUBg{3}$ was chosen because it provides
a clear view of the path $ABA'$, and because these functions are symmetric and
antisymmetric in $\tau_x$, respectively. Note that the two pairs of trajectories
in \Wmnfld{u,S}{UB}and $\tau_x \Wmnfld{u,S}{UB}$ draw together just before $A'$.
Each pair of nearby trajectories emanating from the same {\eqb} splits at
$A'$ and switches allegiance with the pair from the opposite {\eqb}, so
that past $A'$, trajectories on opposite unstable manifolds follow almost identical
paths. The $\tau_x$-antisymmetric long-term behavior of two nearby initial conditions
from $\uUB$ suggests that the path from $B$ to $A'$ is one of weakening $x$ variation,
reaching small but nearly $\tau_x$-antisymmetric $x$ variation near $A'$. After $A'$,
a $\tau_x$-antisymmetric instability comes into play, resulting in long-term 
$\tau_x$-antisymmetric dynamics.

For the parameters of this study, the trajectories investigated so far leave the
region of the \uUB\ and its translations after a few oscillations, so that the
\uUB\ unstable manifold has the characteristics of a chaotic repeller. We expect
that unstable periodic orbits can be found in this region, and we intend to explore
this in a future publication.

\subsection{Transient turbulence}
\label{s:transtTurb}

%%%%%%%%%%%%%%%%%%%%%%%%%%%%%%%%%%%%%%%%%%%%%%%%%%%%%%%%%%%%%%
\begin{figure}
\centering
{\small (a)} \includegraphics[width=4.5in]{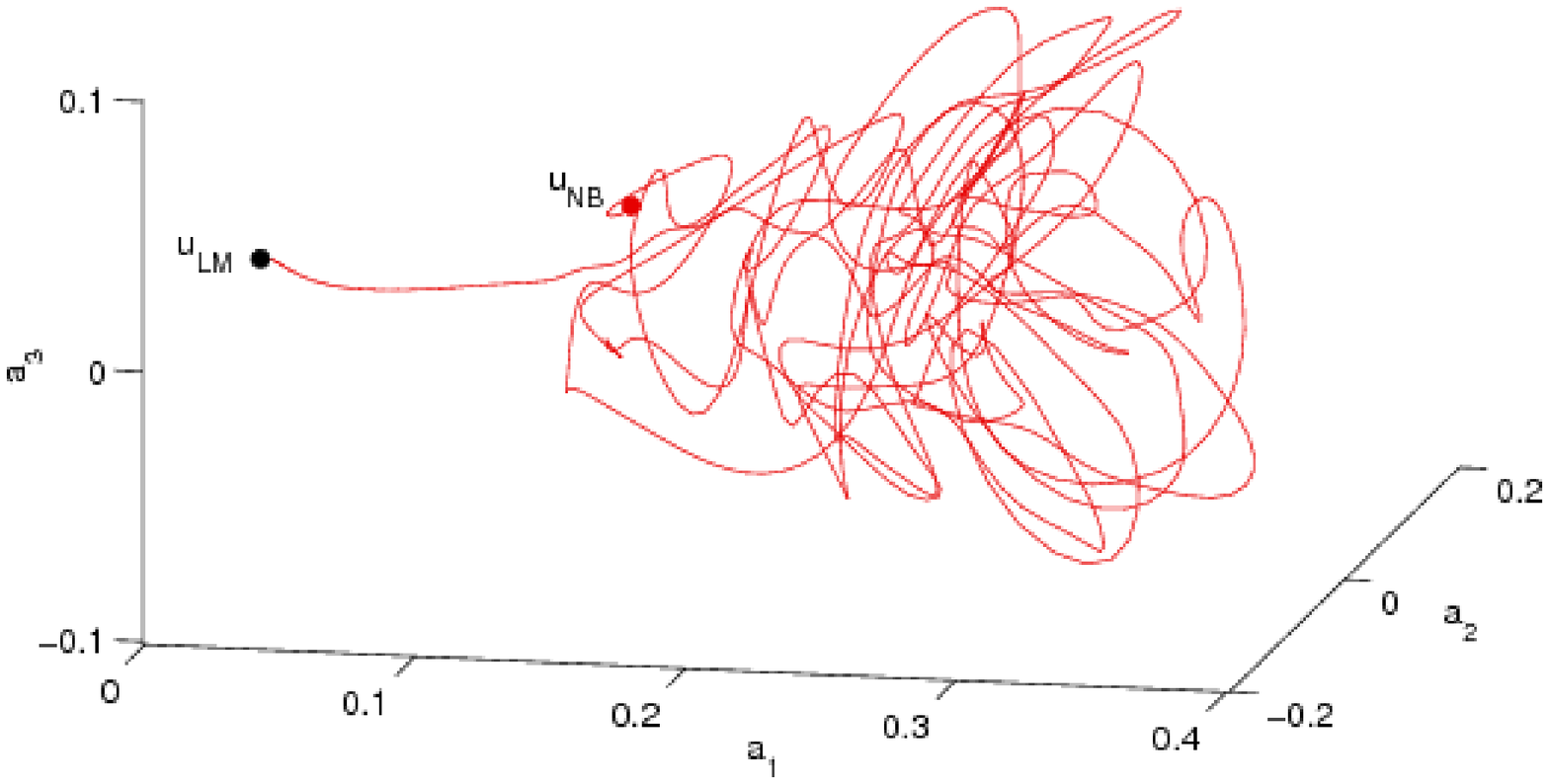} \\
{\small (b)} \includegraphics[width=4.5in]{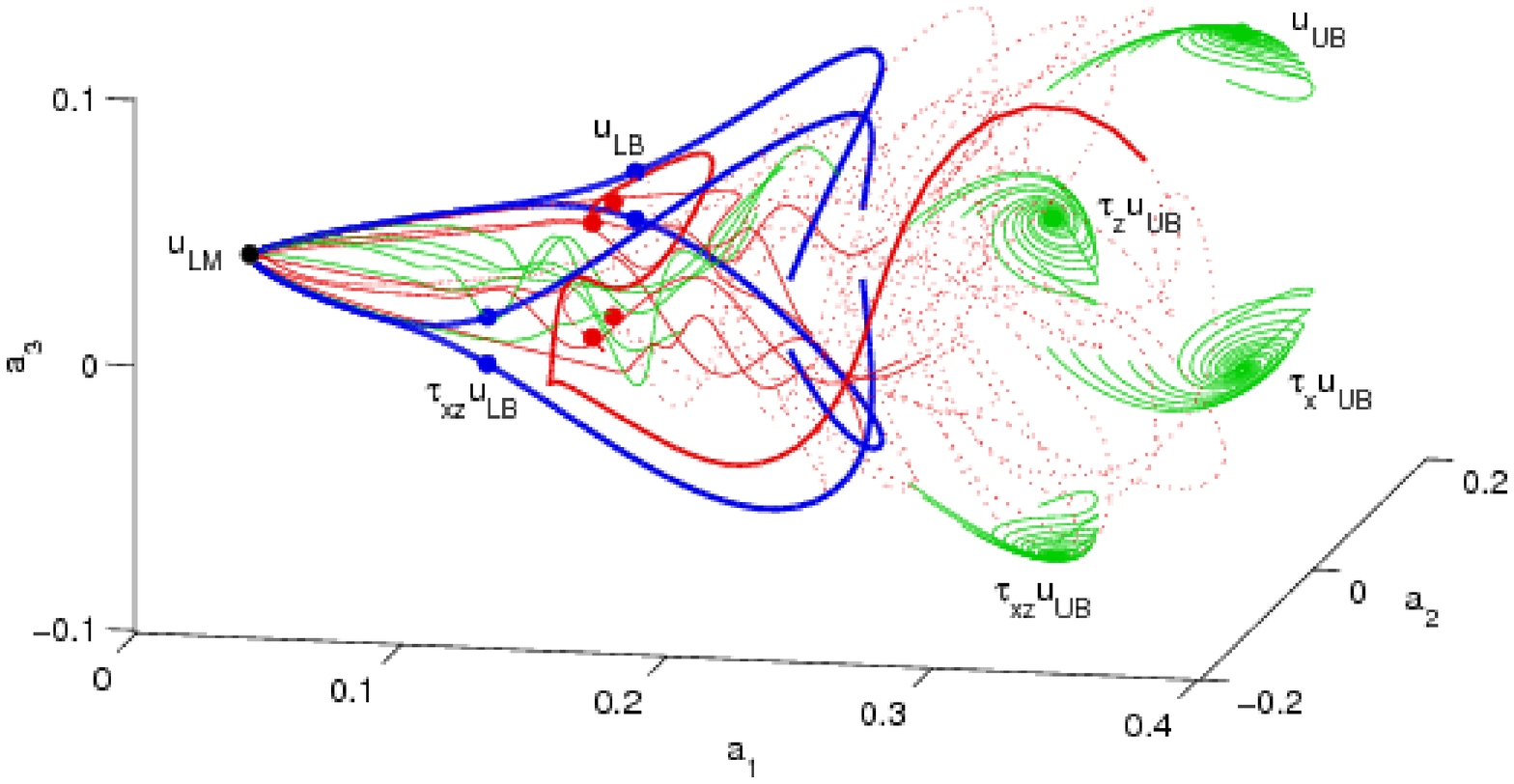}
\caption{
{\bf A transiently turbulent trajectory} in the \uNB unstable
manifold, (a) in isolation (b) in relation to \uLB, \uNB, \uUB, their
half-cell translations, and their unstable manifolds (see
\reffig{f:LBNBUB_portrait_a} and \reffig{f:LBNBUB_unstable_b} for more
detailed labeling of these features). The final decay
to laminar of several other trajectories in the unstable manifolds
of \uNB\ and \uUB\ are also shown. The projection is the same as that
of \reffig{f:LBNBUB_unstable_b}.
        }
\label{f:turbulence}
\end{figure}
%%%%%%%%%%%%%%%%%%%%%%%%%%%%%%%%%%%%%%%%%%%%%%%%%%%%%%%%%%%%%%

The final stop in our stroll through plane Couette {\statesp}
is an illustration of transient turbulence against the backdrop
of the invariant structures featured in previous figures.
For the Reynolds and cell aspect ratios studied here,
all initial conditions investigated so far
ultimately decay to laminar.
\refFig{f:turbulence}(\textit{a}) shows a single trajectory, initiated
as a perturbation of \uNB, that exhibits transient turbulence and then
decays to laminar flow. The coordinate system is \refeq{ek_defn}, the same
as in \reffig{f:LBNBUB_unstable_b}. The region of \statesp\ explored by this
trajectory is typical of all observed transiently turbulent
dynamics in $\bbU_S$.  The trajectory is
unusually long-lived; after leaving the spiralling region around \uNB\ it
wanders for some 1000 nondimensionalized $L/U$ time units before converging on the laminar
state, compared to more typical 200 time-unit lifetimes of other
trajectories initialized as pertubations of \uNB.

When seen in isolation in \reffig{f:turbulence}\textit{(a)},
the turbulent trajectory shows little
discernable order. When plotted within the framework of invariant
structures of the flow, in \reffig{f:turbulence}\textit{(b)},
structure is immediately evident. In this $3d$ perspective, the
decay to laminar flow is confined to a region bounded by the
\uNB\ and \uLB\ unstable manifolds. Transient turbulence occurs
on the far side of laminar from these states, $a_1 > 0.15$,
and in a region shaped roughly by the unstable manifolds of \uLB,
\uUB, \uNB, and their half-box translations. Close inspection
shows that segments of the transient turbulent trajectory follow
the contours of nearby unstable manifolds.

%% file: conclusions.tex
% summary.tex
% $Author: gibson $ $Date: 2008-02-11 00:38:43 +0530 (Mon, 11 Feb 2008) $
% from nsf06am \file{summary.tex  - oct 29 2005 version, printed \today}

% {\bf Intellectual Merit:}
%
Currently a large conceptual gap separates what has been achieved for
low-dimensional dynamical systems and the challenges we face in
understanding infinite-dimensional turbulent flows. Recent computations
of invariant solutions of wall-bounded flows and their agreement with 
the qualitative features of numerical simulations and experiments suggest 
that a dynamical theory of moderate-{\Reynolds} turbulence is within reach.
%Motivated by the recent observations of \recurrStr s in experiments
%and numerical studies, %(reviewed in \refsect{s:ExCohStr}), 
We initiate a systematic exploration of the hierarchy of exact unstable
{invariant} solutions of fully-resolved Navier-Stokes equations in
order to describe the spatio-temporally chaotic dynamics of turbulent
fluid flows in terms of these states.
The key advance reported here is a novel visualization of
moderate-\Reynolds\ fluid dynamics in terms of dynamically invariant,
intrinsic and representation independent coordinate frames. The method
offers an alternative visualization of numerical and/or experimental
data of any dissipative flow close to the onset of turbulence.
In this paper, the visualizations lead to the discovery of a new 
equilibrium solution of plane Couette flow and a heteroclinic connection
between two non-trivial equilibria --to our knowledge the first such 
connection ever observed for Navier-Stokes. We have also computed the 
eigenvalues and symmetries of the three known equilibria of plane Couette 
flow in a small periodic cell with moderate Reynolds number and 
established the low-dimensionality of their unstable manifolds. 

At first glance, turbulent dynamics visualized in \statesp\ might
appear hopelessly complex, but under a detailed examination it appears
much less so than feared: it is pieced together from near visitations
to exact \cohStr s
interspersed by transient interludes. \Eqba, traveling waves, and periodic
solutions of plane Couette flow embody Hopf's vision: a repertoire of recurrent
spatio-temporal patterns explored by turbulent dynamics. We conceive
of turbulence as a walk through a repertoire of unstable recurrent
patterns. As a turbulent flow evolves, every so often we catch a
glimpse of a familiar pattern. For any finite spatial resolution, the
flow approximately follows for a finite time a pattern belonging to a
finite alphabet of admissible fluid states, represented here by a set
of exact \cohStr s.

What new insights does the `unstable \cohStr s program' offer? 
Normal-form models derived from severe truncations of spectral
representations of PDEs - most famously the Lorenz model - capture
\emph{qualitatively} the bifurcations and chaotic dynamics
evocative of those observed in fluid dynamics. In contrast, exact
unstable \cohStr s and periodic orbit theory should provide accurate
\emph{quantitative} predictions for dynamical observables of 
Navier-Stokes (such as the average turbulent drag), for a given flow,
given flow geometry, given \Reynolds\ and other parameters. This
description should lead to quantitative predictions of transport
properties of fluid flows such as bulk flow rate, mean wall drag, and
their fluctuations. The success of computing exact eigenfunctions and 
unstable manifolds also opens a new approach to control of turbulence in
wall-bounded shear flows: perturbations in these directions can be used 
to stabilize or chaperone the flow towards a desired fluid state, and 
not necessarily the laminar one (\cite{KK05,WGW07}).

The \stateDsp\ exploration of \eqba\ and their global unstable
manifolds presented here is the first step. While important in
organizing the turbulent flow, \eqba, being static, do not actually
participate in it. That role is played by the infinity of unstable \po
s densely embedded in the asymptotic attractor. That it is possible to
compute exact $3D$ unstable periodic solutions of Navier-Stokes has
been demonstrated in the pioneering work of \cite{KawKida01}, for \po
s, and \cite{Visw07b}, for \rpo s. However, a combination of novel and
proven numerical and analytical techniques such as variational
solvers, periodic orbit theory, and group representation theory still
needs to be developed in order to systematically explore the hierarchy
of such solutions and to derive the statistics of the flow through periodic 
orbit theory (\cite{DasBuch}).

%% file: ackn.tex
% ackn.tex
% $Author: gibson $ $Date: 2008-02-09 18:19:01 +0530 (Sat, 09 Feb 2008) $

We would like to acknowledge F.\ Waleffe for his very generous guidance 
through the course of this research. We also greatly appreciate 
D.\ Viswanath's guidance in the linearized stability calculations
and his thoughtful comments on drafts. We are very grateful for the 
thoughtful comments of the reviewers. 
P.C.\ and J.F.G.\ thank G.~Robinson,~Jr.\ for support.
J.H.\ thanks R.~Mainieri and T.~Brown, 
Institute for Physical Sciences, for partial support.

%% file: appeTables.tex
% appeTables.tex
% $Author: gibson $ $Date: 2008-02-11 00:38:43 +0530 (Mon, 11 Feb 2008) $

% \section{Tabulation of numerical results}
% \label{s:appeTables}

\begin{table}
\centering
{\small
\begin{tabular}{ccccrr}
$n$ & mode & $k_y$ & $k_z$ &
        Arnoldi $\eigExp[n]_{\text{\tiny LM}}$
          & Analytic $\eigExp[n]_{\text{\tiny LM}}$  \\
\hline
  1,2 & H   & 1 & 0 & -0.00616850 &  -0.00616850 \\
  3,4 & H   & 1 & 1 & -0.02179322 &  -0.02179350 \\
  5,6 & H   & 2 & 0 & -0.02467398 &  -0.02467401 \\
  7,8 & S   & - & 1 & -0.02916371 &  -0.02916371 \\
 9,10 & H   & 2 & 1 & -0.04029896 &  -0.04029901 \\
11,12 & H   & 3 & 0 & -0.05551652 &  -0.05551653 \\
\\
\\
\\
\\
\end{tabular}%
~~~~~%&
\begin{tabular}{c|llc@{}c@{}c}
$n$ & $\eigRe[n]_{\text{\tiny LB}}$~~~  & ~~~$\eigIm[n]_{\text{\tiny LB}}$~~~
                                        & $s_1$&$s_2$&$s_3$ \\
\hline
1       &  0.0501205    &                 & S & S & S \\
2       &  1.878e-06    &                 & - & - & - \\
3       & -1.625e-06    &                 & - & - & - \\
4       & -0.0020054    &                 & A & S & A \\
5       & -0.0065977    &                 & A & A & S \\
6       & -0.0069308    &                 & S & A & A \\
7       & -0.0097953    &                 & S & A & A \\
8       & -0.0135925    &                 & A & S & A \\
9       & -0.0239353    &                 & S & S & S \\
10      & -0.0335130    &                 & S & S & S \\
11      & -0.0370295    &                 & S & A & A \\
12,13   & -0.0454857    & 0.0190660       & A & A & S 
%14,15   & -0.0484668    & 0.1025150       & S & S & S \\
%16,17   & -0.0518223    & 0.0260556       & S & A & A \\
%18      & -0.0554185    &                 & A & S & S \\
%19,20   & -0.0624099    & 0.0311804       & S & S & S
\end{tabular}
}
\caption{
(left) Least stable \ew s of the laminar \eqb\ \uLM\ for $[L_x,L_y,L_z] =
[2\upi/1.14,2,4\upi/5]$ and $\Reynolds = 400$, computed by Arnoldi iteration, compared to
Stokes (S) and heat-equation (H) eigenvalues from analytic formulas. This serves as a test
of accuracy for our {\tt channelflow.org} codes. The heat-equation eigenfunctions have the 
form $\bu(\bx,t) = e^{\lambda t} \sin(\upi k_y y/2) \cos(2\upi k_z z/L_z)) \, {\bf \hat{x}}$ for
$k_y$ even and $e^{\lambda t} \cos(\upi k_y y/2) \cos(2\upi k_z z/L_z)) \, {\bf \hat{x}}$ 
for $k_y$ odd, and eigenvalues $\lambda = -(\upi^2 k_y^2/4 + 4 \upi^2 k_z^2 /L_z^2)/\Reynolds$.
The Stokes eigenvalue listed is the lowest-order mode with $v$ component of the form 
$\hat{v}(y) \cos(2 \upi z/L_z)$, $\hat{v}(y)$ even in $y$ (see \cite{W97}). 
The eigenvalues are ordered in the table by decreasing real part. All laminar 
eigenvalues are real.
(right)
    \uLB\ \eqb\ stability eigenvalues $\lambda = \mu \pm i \omega$ 
    and symmetries of corresponding eigenvectors at same parameter values.
The zero eigenvalues result from the continuous translation symmetry of the flow.
}
\label{t:LBlambda}
\end{table}

\begin{table}
\centering
{\small
\begin{tabular}{c|llc@{}c@{}c}
$n$ & $\eigRe[n]_{\text{\tiny NB}}$~~~  & ~~~$\eigIm[n]_{\text{\tiny NB}}$~~~
                                      & ~~$s_1$&$s_2$&$s_3$ \\
\hline
1       &  0.0306497     &                & A & S & A \\
2,3     &  0.0261952     & 0.056377       & S & S & S \\
4       &  0.0183668     &                & S & S & S \\
5       &  0.0174064     &                & S & A & A \\
6       &  0.0158648     &                & A & A & S \\
7       & -1.047e-07     &                & - & - & - \\
8       & -4.709e-07     &                & - & - & - \\
9       & -0.0045203     &                & A & S & A \\
10      & -0.0048642     &                & S & A & A 
%11,12   & -0.0081398     & 0.254740       & S & S & S \\
%13,14   & -0.0151787     & 0.280906       & A & A & S \\
%15,16   & -0.0195815     & 0.308480       & S & A & A \\
%17      & -0.0233405     &                & A & A & S \\
%18,19   & -0.0280165     & 0.277391       & A & S & A \\
%20      & -0.0359781     &                & S & S & S
\end{tabular}
 ~~~~~ %&
\begin{tabular}{c|llc@{}c@{}c}
$n$ & $\eigRe[n]_{\text{\tiny UB}}$~~~  & ~~~$\eigIm[n]_{\text{\tiny UB}}$~~~
                                      & ~~$s_1$&$s_2$&$s_3$ \\
\hline
1       &  0.0555837    &                 & A & A & S \\
2,3     &  0.0325292    & 0.107043        & S & S & S \\
4,5     &  0.0160591    & 0.039238        & S & A & A \\
6,7     &  0.0152926    & 0.284177        & S & A & A \\
8       &  0.0106036    &                 & A & S & A \\
9       &  1.032e-06    &                 & - & - & - \\
10      &  1.599e-07    &                 & - & - & - \\
11,12   & -0.0141215    & 0.057748        & S & S & S \\
13      & -0.0181827    &                 & S & A & A 
%14,15   & -0.0209193    & 0.173592        & A & A & S 
%16,17   & -0.0242951    & 0.147947        & S & S & S \\
%18,19   & -0.0264681    & 0.312192        & A & S & A \\
%20,21   & -0.0274125    & 0.147147        & A & A & S \\
\end{tabular}
}
\caption{
Stability eigenvalues $\lambda = \mu \pm i \omega$ and symmetries of corresponding eigenvectors: (left) \uNB, 
(right) \uUB\ \eqb\ for $[L_x,L_y,L_z] = [2\upi/1.14,2,4\upi/5]$ and $\Reynolds = 400$.
}
\label{t:NBUBlambda}
\end{table}

Tables \ref{t:LBlambda} and \ref{t:NBUBlambda} list
the least stable linear stability
eigenvalues of the \uLM, \uLB, \uNB, and \uUB\ \eqba, together with
symmetries of corresponding eigenfunctions.
The unstable eigenvalues together with a set of
the least contracting stable eigenvalues are also shown 
in \refFig{f:UBLBNBlambda}.
All numerical results tabulated in this appendix are
computed for \pCf\ with $\Reynolds = 400$ and  
$[L_x,L_y,L_z] = [2\upi/1.14,2,4\upi/5]$. Full sets of
exact invariant solutions, their linear stability eigenvalues and 
eigenfunctions are available on {\tt channelflow.org}, 
(\cite{channelflow}), or can be obtained by a request to authors.